\documentclass[12pt,preprint]{aastex}
\usepackage{graphicx}
\usepackage{lscape}

\slugcomment{Submitted for publication in the Astronomical Journal}

\shorttitle{}
\shortauthors{Allende Prieto et al.}

\begin{document}

\title{The SEGUE Stellar Parameter Pipeline. III. Comparison with
High-Resolution Spectroscopy of SDSS/SEGUE Field Stars\footnote{Based on 
observations obtained with the Hobby-Eberly Telescope 
(a joint project of the University of Texas at Austin, 
the Pennsylvania State University, Stanford University, 
Ludwig-Maximilians-Universit\"at M\"unchen, 
and Georg-August-Universit\"at G\"ottingen), 
the W. M. Keck Observatory 
(operated as a scientific partnership among the California Institute 
of Technology, the University of California and the National Aeronautics and 
Space Administration),
and the Subaru Telescope (operated by the 
National Astronomical Observatory of Japan).}}

\author{Carlos Allende Prieto}
\affil{McDonald Observatory and Department of Astronomy, University of Texas,
    Austin, TX 78712}
\email{callende@astro.as.utexas.edu}    

\author{Thirupathi Sivarani, Timothy C. Beers, Young Sun Lee}
\affil{Department of Physics \& Astronomy, CSCE: 
Center for the Study of Cosmic Evolution, and 
JINA: Joint Institute for Nuclear Astrophysics, Michigan State
University, East Lansing, MI 48824, USA} 
\email{thirupathi, beers, lee@pa.msu.edu}    

\author{Lars Koesterke, Matthew Shetrone, Christopher Sneden, David L. Lambert}
\affil{McDonald Observatory and Department of Astronomy, University of Texas,
    Austin, TX 78712}
\email{lars, shetrone , chris , dll@astro.as.utexas.edu} 

\author{Ronald Wilhelm}
\affil{Department of Physics, Texas Tech University, Lubbock, TX 79409}
\email{ron.wilhelm@ttu.edu}

\author{Constance M. Rockosi, David K. Lai}
\affil{UCO/Lick Observatory, 1156 High Street, Santa Cruz, CA 95064}
\email{crockosi, david@ucolick.org}

\author{Brian Yanny}
\affil{Fermi National Accelerator Laboratory, P.O. Box 500, Batavia, IL 60510}
\email{yanny@fnal.gov}

\author{Inese I. Ivans}
\affil{The Observatories of the Carnegie Institution of Washington, Pasadena, CA; and Princeton 
University Observatory, Princeton, NJ}
\email{iii@ociw.edu}

\author{Jennifer A. Johnson}
\affil{Department of Astronomy, Ohio State University, Columbus, OH}
\email{jaj@astronomy.ohio-state.edu}

\author{Wako Aoki}
\affil{National Astronomical Observatory, Mitaka, Tokyo 181-8588, Japan}
\email{aoki.wako@nao.ac.jp}

\author{Coryn A. L. Bailer-Jones, Paola Re Fiorentin}
\affil{Max-Planck-Institute for Astronomy, K\"onigstuhl 17, D-69117, 
Heidelberg, Germany}

\begin{abstract}

We report high-resolution spectroscopy of 125 field stars previously observed as
part of the Sloan Digital Sky Survey and its program for Galactic studies, the
Sloan Extension for Galactic Understanding and Exploration (SEGUE). These
spectra are used to measure radial velocities and to derive atmospheric
parameters, which we compare with those reported by the SEGUE Stellar Parameter
Pipeline (SSPP). The SSPP obtains estimates of these quantities based on SDSS
$ugriz$ photometry and low-resolution ($R \sim 2000$) spectroscopy. For F- and
G-type stars observed with high signal-to-noise ratios ($S/N$), we empirically
determine the typical random uncertainties in the radial velocities, effective
temperatures, surface gravities, and metallicities delivered by the SSPP to be
2.4 km s$^{-1}$, 130 K (2.2 \%), 0.21 dex, and 0.11 dex, respectively, with
systematic uncertainties of a similar magnitude in the effective temperatures
and metallicities. We estimate random errors for lower $S/N$ spectra based on
numerical simulations.

\end{abstract}


\section{Introduction}

Starting from the sixth public data release (DR-6; Adelman-McCarthy et al.
2007), the Sloan Digital Sky Survey (SDSS) provides estimates of the atmospheric
parameters for a subset of the stars observed spectroscopically in the survey
(those in the approximate range of temperature $4500 \le T_{\rm eff} \le
7500$~K). Following completion of the main survey (SDSS-I), the SDSS
instrumentation has been devoted to several programs, including SEGUE: Sloan
Extension for Galactic Understanding and Exploration, a massive survey of the
stellar content of the Milky Way. Collectively, the suite of computer programs
employed to determine atmospheric parameters from SEGUE data is known as the
SEGUE Stellar Parameter Pipeline (SSPP). Because each of the public data
releases of the SDSS includes and supersedes previous releases, DR-6 also
includes atmospheric parameters for archival stellar observations in SDSS-I.
These stellar parameters are derived by a series of methods, some of which
consider purely spectroscopic information (continuum-normalized spectra), solely
photometry (available in the survey's $ugriz$ system for all targets), or a
combination of photometry and spectroscopy. Paper I in this series describes the
SSPP in detail (Lee et al. 2007a). Paper II compares the predictions of the SSPP
radial velocities and atmospheric parameters with likely members of Galactic
globular and open clusters (Lee et al. 2007b). 

The SDSS uses a CCD camera (Gunn et al. 1998) on a dedicated 2.5m telescope
(Gunn et al. 2006) at Apache Point Observatory, New Mexico, to obtain images in
five broad optical bands ($ugriz$; Fukugita et al.~1996) over approximately
10,000~deg$^2$ of the high Galactic latitude sky. The survey data-processing
software measures the properties of each detected object in the imaging data in
all five bands, and determines and applies both astrometric and photometric
calibrations (Lupton et al. 2001; Pier et al. 2003; Ivezi\'c et al.~2004).
Photometric calibration is provided by simultaneous observations with a 20-inch
telescope at the same site (Hogg et al.~2001; Smith et al.~2002; Stoughton et
al.~2002; Tucker et al.~2006). A technical summary is provided by
York et al. (2000). 

SDSS-I and the ongoing SEGUE survey have already built the largest-ever catalog
of stars in the Milky Way. To date, this includes photometry in five bands for
over 200 million stars and spectroscopy for nearly 300,000 stars
(Adelman-McCarthy et al. 2007). The SDSS spectrographs deliver a resolving power
$\lambda/$FWHM $\sim 2000$ over the wavelength range 380-900 nm. Data reduction
is fully automated, and the SSPP employs the final products from the SDSS
pipeline as input to produce atmospheric parameters (effective temperature,
surface gravity, and metallicity) for stars with
spectral types A, F, G, and K. The best results are obtained for F- and G-type
stars spanning the effective temperature range $5000 < T_{\rm eff}
< 7000$~K.

The quality of the SSPP atmospheric parameters is evaluated using different
approaches, as already described in Paper I: comparing with previously published
spectral libraries, well-studied open and globular clusters, and with
high-resolution observations of field stars. Existing spectral libraries are
useful in order to evaluate and calibrate the SSPP methods that rely on
spectroscopy alone. Allende Prieto et al. (2006) employed the low-resolution
Indo-US library (Valdes et al. 2004), and high-resolution spectra from the
Elodie library (Prugniel \& Soubiran 2001) and the S$^4$N archive (Allende
Prieto et al. 2004). Because the $ugriz$ system was introduced with the SDSS,
the stars included in existing spectral libraries lack photometry in this
system. In addition, these are relatively bright stars, typically with $V<14$
mag, brighter than the bright magnitude limit of the SDSS imaging. The bright
magnitude limit for the SDSS is set by the saturation threshold of the detectors
at the sidereal driftscan rate of the survey. Obtaining data for these brighter
stars would require special-purpose observations with a very different
instrument configuration, which would call into question their value as
calibration observations for the otherwise homogeneous imaging survey.

Star clusters provide stringent tests of the SSPP, as the same metallicity
should be derived for stars that explore wide ranges of masses and luminosities.
Paper II in this series examines SSPP results for likely members of clusters
included in DR-6. One cannot choose clusters with any given metallicity, but has
to take what is provided by nature and accessible from Apache Point.
Furthermore, the effective temperatures and surface gravities for the members of
any given cluster are very strongly correlated, depending on age and chemical
composition. This leads to a patchy coverage of the parameter space. Field stars,
on the other hand, can be chosen to provide better coverage 
and, therefore, naturally complement the clusters. Among the stars
spectroscopically observed with SDSS, those in the range $14 < V
<16.5$ mag can be observed at high spectral resolution with large-aperture telescopes 
and modest integration times. Due to the vast size of the SDSS stellar
sample, these stars can be selected to more uniformly cover the parameter space
of stellar properties, and have the additional benefit that photometry is
already available for them in the SDSS native system.

This paper, the third in the SSPP series, is devoted to the analysis of 125 SDSS
stars newly observed at high-resolution with the Hobby-Eberly, Keck, and Subaru
telescopes. Section 2 describes the sample selection and the observations. The
determination of radial velocities and atmospheric parameters, based on these
observations, are discussed in \S 3 and \S 4, respectively. Section 5 describes
the results for several well-known standard stars observed with the Hobby-Eberly
Telescope. Section 6 compares the parameters derived from high-resolution
spectroscopy with those from the SSPP. Section 7 describes numerical experiments
that explore how the parameters degrade at lower signal-to-noise ratios. Our
conclusions are summarized in \S 8.

\section{Observations}

The majority of the data presented in this paper were obtained with the
Hobby-Eberly Telescope (HET; Ramsey et al. 1998), located in West Texas, making
use of its High Resolution Spectrograph (HRS; Tull 1998). Additional spectra
were obtained with the Keck Observatory, using both the High Resolution Echelle
Spectrometer (HIRES; Vogt et al. 1994) and the Echelle Spectrograph and Imager
(ESI; Sheinis et al. 2002), and with the Subaru telescope and the
High Dispersion Spectrograph (HDS; Noguchi et al. 2002), both located on Mauna
Kea, Hawaii. Table \ref{table1} summarizes the basic information concerning the
spectroscopic observations; more details are provided below.

\subsection{Sample selection}

Field stars with previous spectroscopic observations from SDSS-I or SEGUE were
selected for follow-up spectroscopy at higher resolution. Based on preliminary
SSPP atmospheric parameters, targets were initially chosen to span the range
$5200 < T_{\rm eff} < 7000 $ K, $1.5 < \log g < 5.5$, and $-2.5<$ [Fe/H]$<
0.5$\footnote{Here and throughout the paper we equate metallicity with iron
abundance, and use the notation [Fe/H]$\equiv \log \left( \frac{\rm N(Fe)}{\rm
N(H)} \right) - 
\left( \frac{\rm N(Fe)}{\rm N(H)} \right)_{\odot}$, where N represents the number
density of atoms.}. Our targets are relatively bright; most satisfy $g < 15.5$
mag. In addition, a number of cooler red giants were also included in the
sample, expanding the initial range of temperatures.

\begin{figure}[t!]
\centering
\includegraphics[width=12.cm,angle=0]{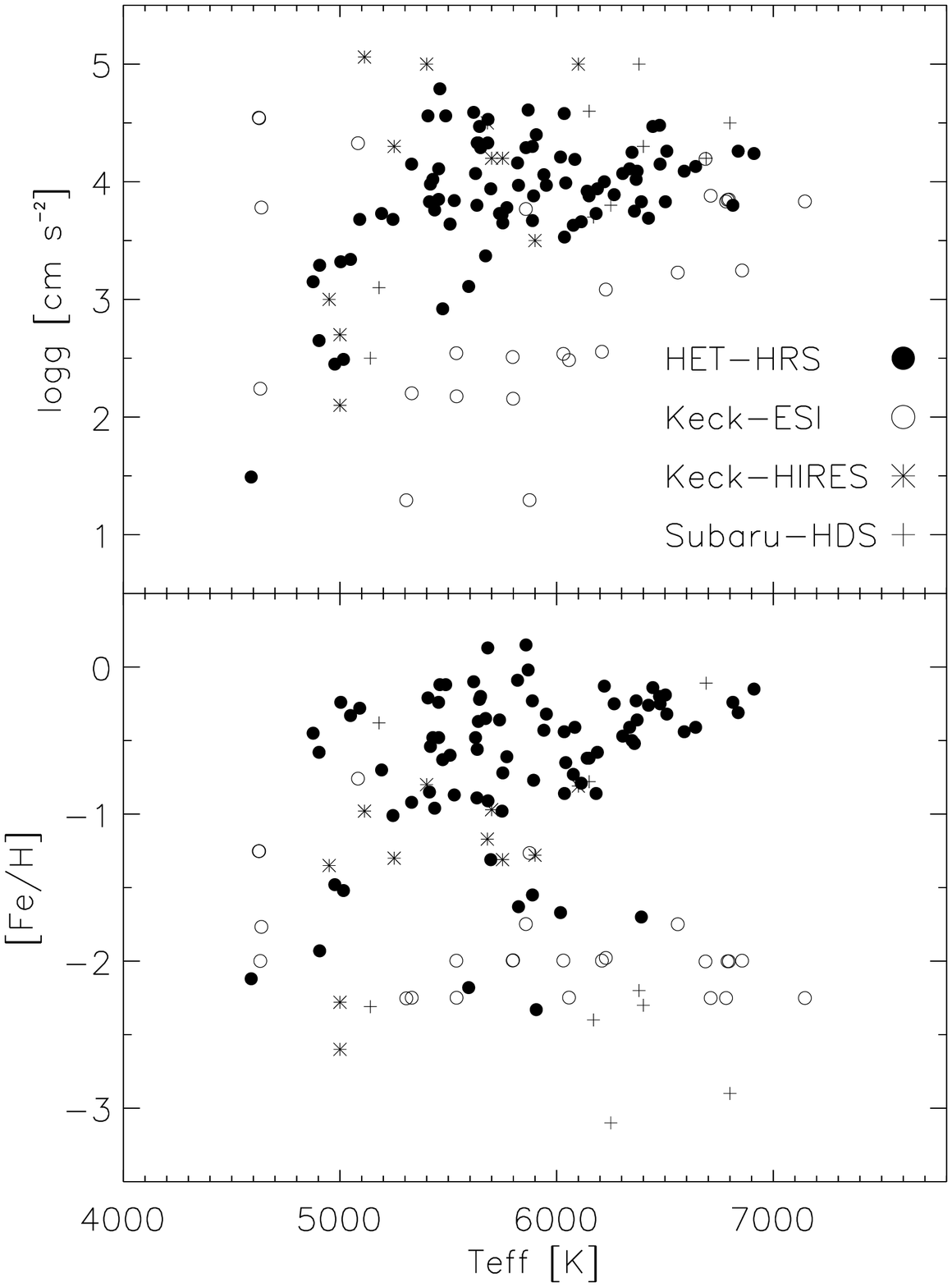}
\protect\caption[ ]{
Distribution of the sample of SDSS/SEGUE stars with available high-resolution
spectra over the parameter space considered herein.
}
\label{sample}
\end{figure}

Figure \ref{sample} illustrates the coverage of parameter space occupied by our
targets. Some 300 stars were placed in the HET queue between November 2005 and
October 2006, despite the fact that time was only allocated for observations of
about 100 of them. This over-booking strategy allows for very efficient use of
the HET queue schedule (Shetrone et al. 2007). The time on Keck and Subaru was
used mainly to increase the target density at low metallicities and cooler
temperatures.

\subsection{HET spectra}

On the HET, a 316 grooves mm$^{-1}$ cross-dispersing grating, and a $2"$-wide 
slit collecting 80\% of the light from the $3"$-diameter science fibers, were
chosen to provide nearly full spectral coverage between 400 and 800 nm at a
resolving power $R=\lambda/{\rm FWHM} \simeq 15000$. Some 280 spectra of 115
stars were obtained. The observations were scheduled at low priority on the HET
queue, and most were obtained during bright time. Below we discuss only the 81
stars that appeared single-lined, did not exhibit the characteristics broad
lines, and had at least one spectrum with no obvious signs of background light
(since no sky fibers were used), and a $S/N$ per pixel at 520 nm
in excess of 20/1.

Data reduction was performed independently at the University of Texas and at
Michigan State University (MSU). The reduction at Texas was done automatically,
with a pipeline based on IRAF\footnote{IRAF is distributed by the National
Optical Astronomy Observatories, which is operated by the Association of
Universities for Research in Astronomy, Inc. under cooperative agreement with
the National Science Foundation.} scripts, while a more interactive procedure,
also based on IRAF packages, was employed at MSU. Both reductions included bias
removal and flatfield correction, but the former corrected for scattered light
with the task {\tt apscatter}, while the latter removed the background for each
order from neighboring areas. The results are generally in excellent agreement.
Multiple observations were typically obtained for each object. With the
exception of nine stars with the lowest $S/N$, individual exposures were
analyzed independently, and the derived atmospheric parameters averaged.

\subsection{Keck-HIRES spectra}

Fourteen objects were observed with the red configuration of the Keck I
High Resolution Echelle Spectrometer (HIRES) and new 3-chip CCD
mosaic, with an on-chip binning of $1\times2$. The C1 decker, which
has a $7.0 \times 0.861"$ slit, was used. This setting yields a
resolving power of $R \sim 40000$. 
The spectra cover a wavelength range of 414--849 nm.
Most of the objects had more than two exposures, and exposure times of
300--1500 sec. The data were reduced at Carnegie Observatories,  
using version 4.0.1. of the  MAuna Kea Echelle Extraction 
data reduction package (MAKEE\footnote{MAKEE was developed by T. A. 
Barlow for the reduction of Keck I HIRES data taken with the 
new 3-chip CCD mosaic. It is freely available from the Keck Observatory}).  
The final $S/N$ per pixel was approximately 80/1 at 520 nm.

\subsection{Keck-ESI spectra}

The Keck~II ESI spectrograph was used in the echellete mode.
Twenty seven objects were observed  with exposure times 
ranging from 300 to 1200 sec. 
The resolving power is approximately 7000, using
a slit width of 0.74\arcsec. The wavelength coverage is 390-1100 nm. 
Data reduction was performed at Santa Cruz 
using IRAF scripts (see Lai et al. 2004).
The $S/N$ per pixel was in the range 30/1--60/1 at 520 nm.

\subsection{Subaru spectra}

The Subaru HDS was used to observe nine of our program objects with a resolving
power of $R \sim 45000$, covering 300-580 nm. The blue cross disperser was chosen for the
observations, with 400 grooves mm$^{-1}$ and blaze angle of 4.76\degr.
Most of the objects had only one exposure.
Standard data reduction procedures (bias subtraction, flat-fielding, background
subtraction, extraction, and wavelength calibration) were carried out with the
IRAF echelle package.  Suspected cosmic-ray hits are removed using the technique
described by Aoki et al. (2005). The $S/N$ per pixel was roughly 80/1 at 520 nm.

\section{Radial Velocities}

Following the same strategy as for the data reduction, the radial velocities for
HET spectra were measured independently at Texas and MSU by three different
methods. There were 10 observations of four radial velocity standards, which are
discussed in Section \ref{standards}.

For the Keck-ESI and the MSU reductions of the HET spectra, radial velocities
were derived from cross correlations with the solar spectrum between 480 and 530
nm (Wallace, Hinkle, \& Livingston 1998). After the spectra were analyzed and the
atmospheric parameters determined, as explained below (\S \ref{sivapar}), the
cross-correlation was repeated using the best-fitting models as templates.
Heliocentric corrections were estimated using the IRAF task {\it rvcor}. The
radial velocities for the Keck-HIRES data were estimated by cross correlation
using the positions of about 100 Fe I and 10 Fe II lines. Heliocentric
corrections were already applied during data reduction by the MAKEE package.

Radial velocities were derived for the Texas-reduced HET spectra by measuring
the central wavelengths of several hundred Fe I lines and comparing to
laboratory values (Nave et al. 1994). The distribution was then fit by a
Gaussian plus a background parabola, in order to determine the mean, the sigma,
and the error of the mean. The heliocentric correction was estimated with the
IRAF task {\it rvcor}, then applied in order to obtain the final radial
velocity.

The Texas-reduced HET spectra were also cross-correlated with a library of
synthetic spectra smoothed to the appropriate resolution, in order to measure
the Doppler shifts. The library covers a region of 4 nm around H$\beta$, and
samples uniformly in $T_{\rm eff}$ the spectral types F to mid-K (4500 to
7500~K), with surface gravities $1.0<\log g<5.0$, and metallicities $-2.5 < $
[Fe/H] $ <0.5$. Each synthetic spectrum was cross-correlated with each HET
spectrum, and the peak of the cross-correlation was fit with a Gaussian using
the IDL routine {\it xc} (Allende Prieto 2007). The Doppler shift is estimated
as the mean value for the 10\% of the synthetic spectra that best fit the
observed spectrum. The heliocentric correction was computed with the routine
{\it baryvel} (see Stumpff 1980) from the IDL astro library\footnote{See http:
//idlastro.gsfc.nasa.gov/}, and applied. We note that heliocentric corrections
derived in this manner differed by those from IRAF's rvcor task by no more than
0.2 km s$^{-1}$.

In summary, three different procedures for radial velocity estimation were
applied to the HET spectra: (1) cross-correlation with the solar spectrum in the
480--530 nm region, (2) direct measurement of the wavelength shifts of atomic
iron lines, and (3) cross-correlation with a library of synthetic spectra in the
vicinity of H$\beta$. Cross-correlation with the solar spectrum was the only
method applied to the Subaru and Keck spectra. This method and the Fe I method
agree with one another slightly better than with the third technique (for HET
stars): excluding the spectra of SDSS J033530.56-010038 
and SDSS J074151.21+275319,  
we find an rms scatter of 1.6 km s$^{-1}$. 
Thus, we adopt the average of these two methods for
all HET stars and exclude these two stars in the comparison with the radial
velocities from the SSPP. The radial velocities for the HET stars are listed in
Table \ref{params}; those for the rest of the sample are listed in Table
\ref{params2}.

\section{Analysis}

The majority of our program stars were observed with HET-HRS using a single 
setting, but the rest of the spectra from Keck and Subaru fill important gaps in
the parameter space. The HET data were analyzed by an automated spectral fitting
technique at the University of Texas. The rest of the spectra were analyzed by a
second method for automated spectral fitting (Keck-ESI), or by more traditional
methods, using line equivalent widths (Subaru-HDS and Keck-HIRES) at MSU. In
order to take advantage of both the homogeneity of the HET spectra, and the
expanded coverage of the rest of the observations, we separately consider these
two data sets, designated below as ``HET'' and ``OTHERS''. One star,
SDSS  J180922.45+223712, 
was observed both with HET and Subaru.  

\subsection{HET analysis}
\label{carlospar}

The determination of atmospheric parameters for HET spectra at Texas was based
on fitting the spectroscopic observations in the range 500-521 nm. This region includes many
individual lines, but it is dominated by transitions of neutral iron, calcium
and magnesium. The spectra were continuum-normalized. The search for the optimal
solution is based on the Nelder-Mead algorithm (Nelder \& Mead 1965), with model
spectra interpolated using a third-order Bezier scheme, but otherwise the same
code and strategy described by Allende Prieto et al. (2006) is used. The code is also
the same used by the SSPP for the methods described in \S 4.1 of Paper I. The
main difference between the {\tt ki13} grid used in the SSPP and the one
employed here is the spectral resolution, which is now $R=7700$, instead of
$R=1000$. With only three fitting parameters (effective temperature, surface
gravity and overall metal abundance), a scaled solar composition is implicit in
the analysis, considering an enhancement of the $\alpha$ elements for [Fe/H]$ <
0$. Note that the same Nelder-Mead algorithm, but a different implementation, is used
for the analysis of the Keck-ESI data at MSU, as described below.

\begin{figure*}[ht!]
\centering
\includegraphics[width=11.cm,angle=90]{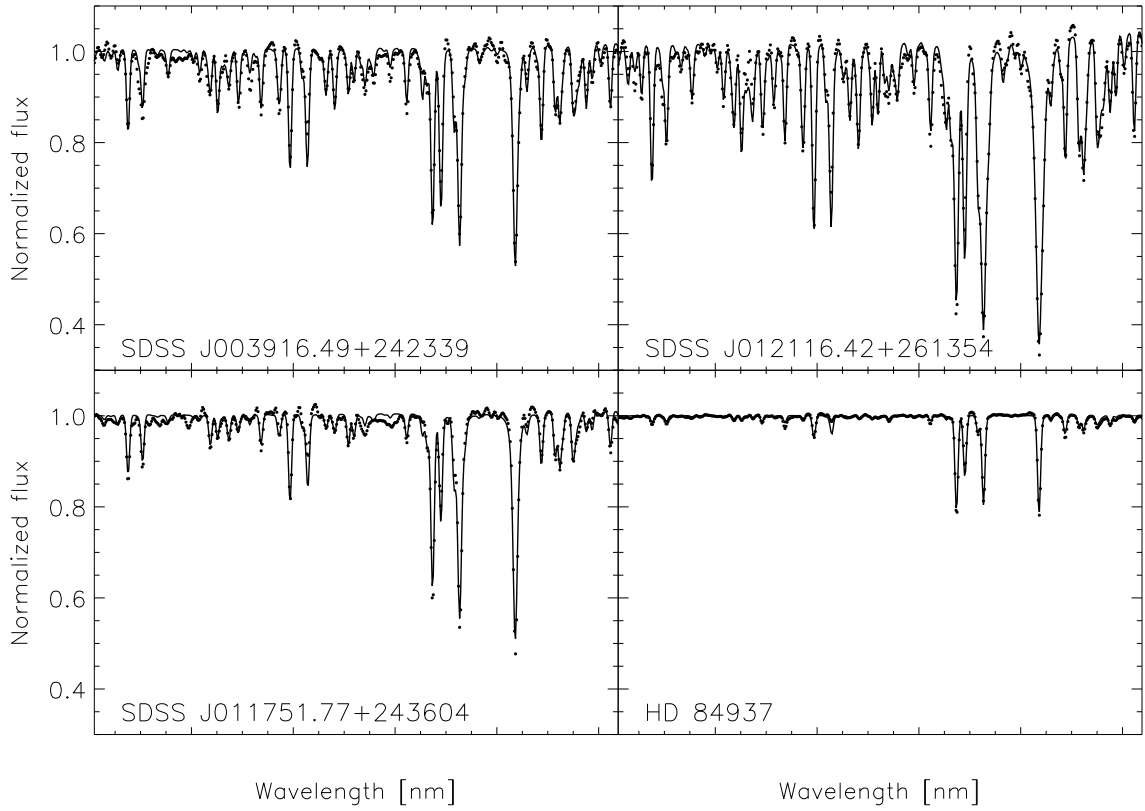}
\protect\caption[ ]{
Fittings to individual HET observations for
three of the program stars, and the metal-poor standard star HD~84937. The 
dots correspond to the observations; the solid lines identify the
best-fitting models, obtained from cubic Bezier interpolation in the original
grid.}
\label{fits}
\end{figure*}

It should be emphasized that although the HET-HRS spectra have a resolving power
of $R=15000$, the analysis is performed at a lower resolution. By smoothing both
the observed and the synthetic spectra to $R=7700$, we effectively eliminate the
effects of stellar rotation, and potential variations with time in the PSF of
the spectrograph, increasing the original signal-to-noise ratio per pixel and
speeding up the calculations. The sacrifice in resolution has a negligible
impact on the final accuracy of the derived atmospheric parameters, as checked
from the analysis of several hundred spectra from the Elodie library at both
$R=15,000$ and $R=7700$. Figure \ref{fits} illustrates the fits for three
program stars and for the metal-poor standard HD~84937, all observed on the HET.
The internal consistency of the derived atmospheric parameters for different
observations of the same target is excellent, typically $\sigma=$ 32 K, 0.05
dex, and 0.02 dex for $T_{\rm eff}$, $\log g$, and [Fe/H], respectively.

The analysis is simplified by assuming a relationship between the abundance
ratio of the alpha elements to iron and the iron abundance (Beers et al. 1999;
Eq. 2 in Allende Prieto et al. 2006), but it is well known that such a
relationship does not apply to all stars in the Galaxy. For example, Reddy et
al. (2006) find different slopes for the change in [$\alpha$/Fe] with [Fe/H] for
stars in the thin- and thick-disk populations. The halo values are most likely
similar to those for the thick disk. Using the average of [Mg/Fe], [Si/Fe],
[Ca/Fe], and [Ti/Fe], Reddy et al. find that approximately linear trends apply,
although they differ somewhat from the relationship adopted in our calculations.
Inspection of their fits suggest slopes of $-0.14$ dex/dex and $-0.07$ dex/dex,
and intercepts at [Fe/H]$=0$ of $+0.00$ and $+0.17$, for the thin- and
thick-disk populations, respectively. Oxygen may not follow the same behavior
(Ram\'{\i}rez et al. 2007), as it appears to exhibit a more pronounced slope for
thin-disk stars, but Mg and Ca are the relevant elements for the spectral window
we are using. In any case, the use of a single relationship for all of the alpha
elements is only an approximation.

Our adopted relationship predicts [$\alpha$/Fe]$= +0.27$, +0.13, and +0.00 at
[Fe/H]$=-1.0$, $-0.5$, and 0.0, respectively, while the results of Reddy et al.
indicate [$\alpha$/Fe]$= +0.14$, +0.07, and 0.00 for the thin-disk population,
and [$\alpha$/Fe]$= +0.24$, +0.21, and +0.17 for the thick-disk population,
respectively, at the same metallicities. Halo stars exhibit similar
[$\alpha$/Fe] ratios as thick-disk stars with [Fe/H]$< -0.7$. These differences
have only a small impact on our results. The parameters for thin-disk stars with
[Fe/H]$\sim -1$ (provided they exist), or for thick-disk stars with solar
metallicity (provided they exist), would have a maximum systematic error of 0.2
dex in surface gravity and metallicity, and 100~K in $T_{\rm eff}$. At the
intermediate metallicities where the two populations overlap, errors would
amount to about half of the maximum values.

\begin{figure*}[ht!]
\centering
\includegraphics[width=8.cm,angle=0]{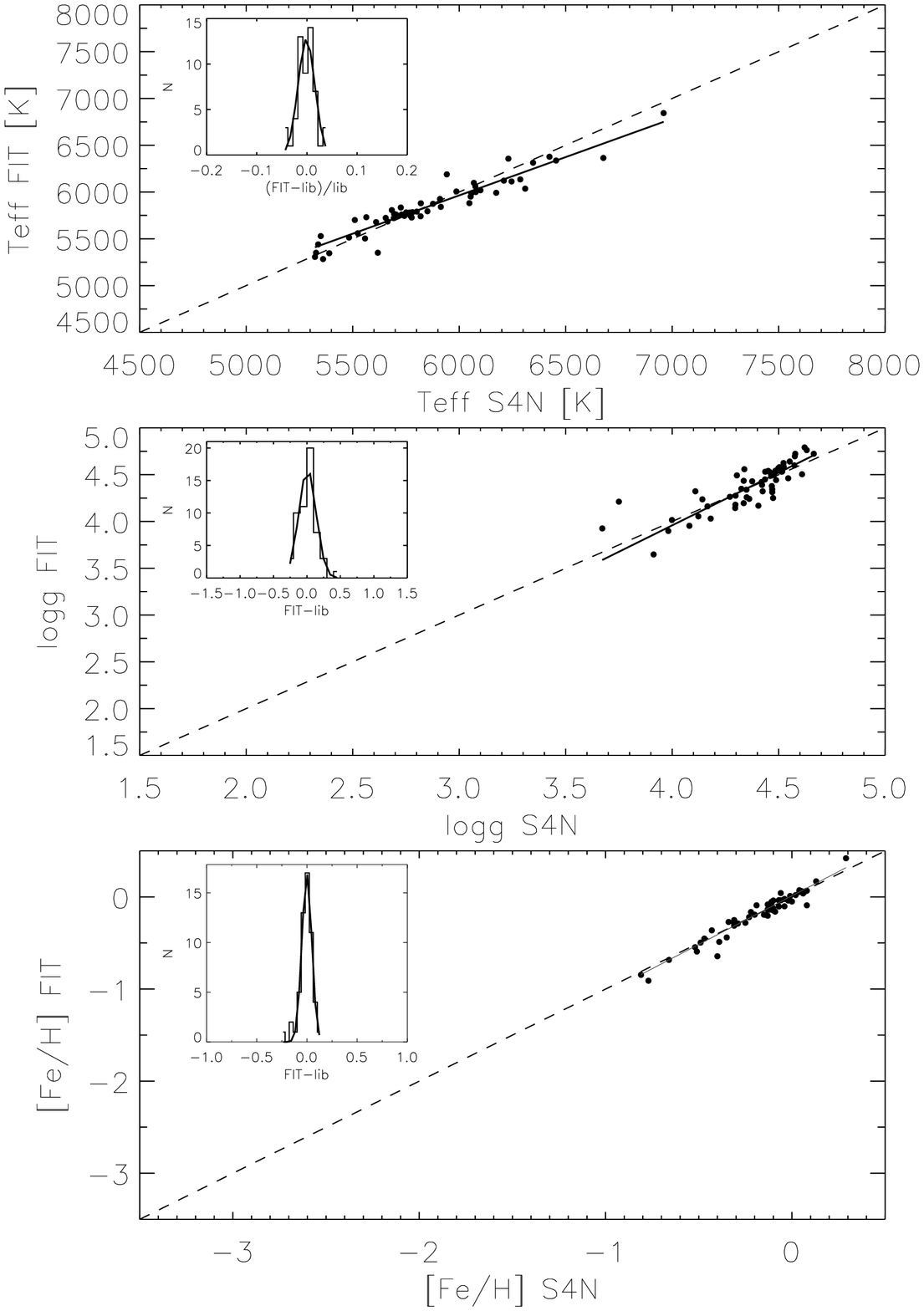}
\includegraphics[width=8.cm,angle=0]{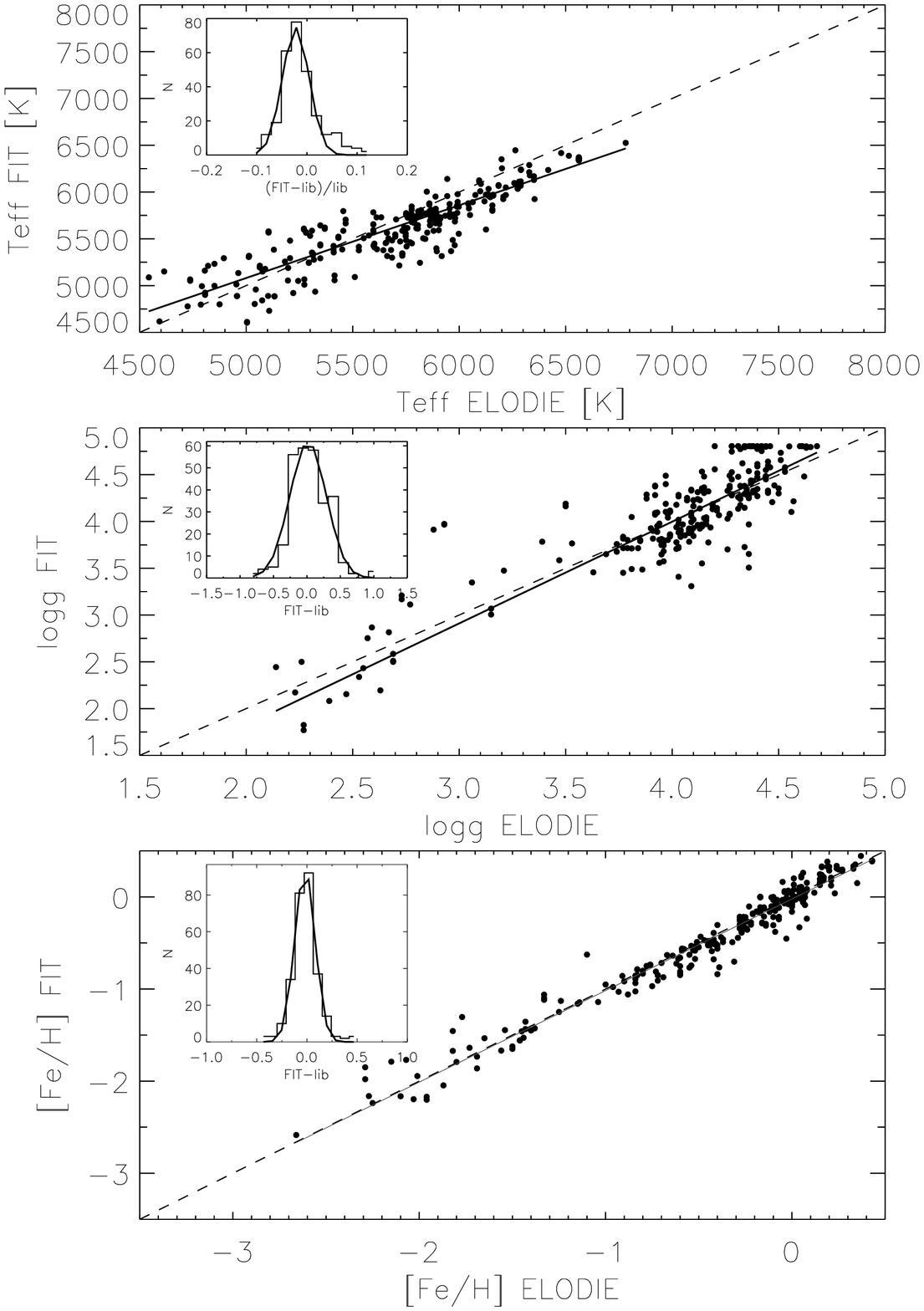}
\protect\caption[ ]{
Comparison between our derived parameters for the spectra in the
S$^4$N and the Elodie.3 libraries with those in the catalogs associated 
with the libraries (see \S \ref{carlospar}). 
The dashed lines indicate a slope of unity; the
solid lines are linear fits to the data. The insets show the
differences between the fit and catalog parameters, as well as Gaussian
models employed to make robust estimates of the median and standard
deviation, as shown in Table \ref{libraries}.
}
\label{figcor}
\end{figure*}

The analysis procedure was tested and calibrated using two spectral libraries
from the literature: S$^4$N (Allende Prieto et al. 2004), and the Elodie.3
library (Prugniel \& Soubiran 2001). Our comparison is limited to stars in these
libraries with effective temperatures between $4500<T_{\rm eff}<7000$ K, and, in
the case of the Elodie library, with reliable parameters ($Q_{Teff} \ge 2$,
$Q_{\log g} \ge 1$, and $Q_{\rm [Fe/H]} \ge 3$, where $Q$ represents {\it
reliability} as defined by the Elodie team). We estimate random and systematic
uncertainties by fitting Gaussian models to the differences between the
parameters derived for the spectra in these libraries, and their associated
catalogs. Our results are systematically different from the S$^4$N catalog
parameters by $+5$\% in $T_{\rm eff}$, +0.20 dex in $\log g$, and $-0.23$ dex in
[Fe/H]. After correcting for these zero-point offsets, the differences between
our parameters and those in the libraries' catalogs are illustrated in Figure
\ref{figcor}; statistics are presented in Table \ref{libraries}, where
the $\sigma_{\rm rms}$ is derived from Gaussian fittings. 

The larger scatter found for the Elodie library is expected, since the
corresponding catalog values do not have a homogeneous source, but are mostly
compiled from the literature. In addition, the quality of the original spectra
in this library is lower than those in the S$^4$N library. The $1\sigma$
uncertainties derived from the comparison with the S$^4$N library are adopted as
external errors, and added in quadrature to the internal estimates.

The empirically determined corrections from the S$^4$N library for surface
gravity and metallicity work as well for the Elodie library. While the first
library is dominated by spectra of thin-disk stars, the second balances
thin-disk, thick-disk, and halo populations, spanning metallicities between
$-3.0$ and $+0.5$. With the zero points determined from the comparison with the
S$^4$N library, our effective temperatures are roughly 2\% lower than those in
the Elodie library. This difference is expected, since the temperatures in the
S$^4$N catalog were obtained from the infrared flux method (IRFM) calibrations
of Alonso et al. (1996, 1999), while most of the values reported in the Elodie
catalog are from spectroscopic analyses. It is well known that the spectroscopic
(excitation balance of neutral iron lines, as described in
\S \ref{sivapar}) temperature scale is about 150~K warmer than the IRFM scale for
these spectral types (see, e.g., Heiter \& Luck 2003, Yong et al. 2004). For
consistency with the results for the OTHERS sample, described below, the warmer
(Elodie) temperature scale is adopted.

\subsection{OTHERS analysis}
\label{sivapar}

The atmospheric parameters for the Keck-ESI spectra were derived at MSU, using a
grid of synthetic spectra and the IDL optimization routine AMOEBA (see Press et
al. 1986), which also employs the Nelder-Mead algorithm.

A total of 13662 synthetic spectra were generated with a sampling step of
$\delta\lambda = 5 \times 10^{-4}$ nm, covering the wavelength range 480--530
nm. The parameter space spans the range 3500 to 9750~K in $T_{\rm eff}$, 0.0 to
5.0 in $\log g$, and $-2.5$ to 0.0 in [Fe/H], for $\xi =$ 1, 2, 3 km s$^{-1}$.
The stellar model atmospheres used for the synthetic spectra are the NEWODF
models by Castelli \& Kurucz (2003), which include updated opacities for TiO
(Schwenke 1998) and H$_{2}$O (Partridge \& Schwenke 1997). The NEWODF models use
solar abundances by Grevesse \& Sauval (1998) and no convective overshooting
(Castelli et al. 1997). The synthetic spectra are generated using the {\tt
turbospectrum} synthesis code (Alvarez \& Plez 1998), and employ recent
calculations of the broadening of Balmer lines (Barklem et al. 2000), and strong
metallic lines (Barklem \& Aspelund-Johansson 2005 and references therein) by
collisions with hydrogen atoms. The linelists employed come from a variety of
sources. Atomic line data are taken mainly from the VALD compilation (Kupka et
al. 1999) as of 2002, and in some cases updated from the literature. The atomic
linelist also includes hyperfine splitting for interesting lines. Linelists for
the molecular species CH, CN, TiO, CaH and OH were provided by B. Plez (see Plez
1998; Plez \& Cohen 2005), while the data for the NH, C$_{2}$ and MgH molecules
are from Kurucz (see http: //kurucz.harvard.edu/LINELISTS/LINESMOL/). The
solar abundances compiled by Asplund, Grevesse \& Sauval (2005) were adopted.
Finally, the synthetic spectra were reduced to a resolution of $R = 7000$ by
convolving with a Gaussian. The SSPP parameters were supplied as initial
guesses.

The analysis of the Keck-HIRES and Subaru-HDS data was performed at MSU using
the equivalent widths of Fe I and Fe II lines to constrain $T_{\rm eff}$, $\log
g$, [Fe/H], and the microturbulence. The $T_{\rm eff}$ is determined from the
excitation equilibrium of Fe I lines, by forcing a null trend in the excitation
potential versus Fe I abundance. The $\log g$ is determined from the ionization
equilibrium of Fe I and Fe II lines. The microturbulence is estimated by forcing
a null trend in the equivalent width versus abundance relation. In our analysis
we used only lines with equivalent widths $\le 120$m\AA\ , so as to avoid the
non-linear part of the curve of growth. The atomic data for the Fe I and Fe II
lines are from the VALD compilation, and from fits to the solar spectrum. We
also checked our estimations by fitting the Balmer line profiles. We have
removed three objects from the Keck-HIRES sample; two of them exhibited very
broad lines, apparently due to rapid rotation, while one object was a
double-lined spectroscopic binary. For one star, SDSS J205025.83-011103.8, the
SSPP did not return measurements.

\section{Standard Stars}
\label{standards}

The HET sample contains four well-known radial velocity standard stars that have
multiple and recent high-resolution analyses in the literature. The stars
HD~8648 and HD~84737 have been reported by Nidever et al. (2002) as constant in
radial velocity to better than 0.1 km s$^{-1}$ over several years; their
heliocentric radial velocities are 0.92 and 4.90 km s$^{-1}$, respectively.
Nordstr\"om et al. (2004) provide values consistent with these measurements. The
radial velocity of HD~71148 has been measured by Nordstr\"om et al. as $-32.6
\pm 0.1$ km s$^{-1}$, with consistent measurements reported by Barnes, Moffett
\& Slovak (1986). Nordstr\"om et al. also included HD~84937 in their sample, with a
radial velocity of $-14.5 \pm 0.2$ km s$^{-1}$, in good agreement with previous
data from Carney et al. (2001).

The average velocities measured from the HET spectra of HD~8648 (5
observations), HD~84737 (1 observation), HD~71148 (2 observations), and HD~84937
(2 observations) are $0.34$, $4.03$, $-33.39$, and $-12.31$ km s$^{-1}$,
respectively. This indicates that a negligible offset exists between the HET
values and those adopted from the literature: $-0.01 \pm 0.74$ km s$^{-1}$, with
an rms scatter of 1.47 km s$^{-1}$.

HD~8648 has been spectroscopically studied by Mishenina et al. (2004) and
Valenti \& Fisher (2005). HD~71148 has been analyzed by Fuhrmann (2004), Lambert
\& Reddy (2004), Mishenina et al. (2004), and Valenti \& Fisher (2005). HD~84737
was observed by Chen et al. (2002), Luck \& Heiter (2006), and Valenti \&
Fisher (2005). Finally, the spectrum of the halo subdwarf HD~84937 has been
analyzed, among others, by Korn, Shi, \& Gehren (2003), Nissen et al. (2007), and
Ryde \& Lambert (2004). The agreement among these studies on the atmospheric
parameters for each star is excellent -- the rms scatter is less than 80~K for
$T_{\rm eff}$, 0.1 dex for $\log g$, and 0.05 dex for [Fe/H]. We adopt 
average values of these parameters for our analysis.

A comparison between our adopted literature parameters and those derived from
our own analysis of high-resolution HET spectra is provided in Table \ref{stds}.
The effective temperatures of these stars span a limited range, as do
their surface gravities, but these objects provide one way to assess the
adopted zero points for our atmospheric parameters. On average, our temperatures
are 18~K warmer, our gravities $-0.05$ dex lower, and our metallicities
$-0.02$~dex lower than the average literature values. These tiny differences
indicate no detectable offsets in our derived atmospheric parameters. The rms
scatter between our parameters and the literature values are 96~K (2 \%),
0.15~dex, and 0.04~dex, in $T_{\rm eff}$, $\log g$, and [Fe/H], respectively.
These estimates are also in excellent agreement with the results based on
comparison with the S$^4$N library shown in Table \ref{libraries}. Most of our
standard stars have near solar metallicity; the same applies to the stars in
the S$^4$N library.  However, the agreement with the literature values for HD~84937,
at [Fe/H] $\simeq -2.1$, for $T_{\rm eff}$ and [Fe/H], does not seem to degrade
significantly. The surface gravity, on the other hand, does exhibit a larger
difference, of about 0.2~dex, which suggests a lower precision for this
parameter at low metallicity, at least for the HET spectra.


\section{Comparison With SSPP Estimates}

\subsection{Radial velocities}

Our two preferred radial velocity determinations for the HET spectra agree with
one another with an rms scatter of 1.6 km s$^{-1}$ (average difference of 0.9 km
s$^{-1})$. This value is consistent with the scatter inferred for the four
radial velocity standards, as described in \S \ref{standards}. The radial
velocities measured in the SDSS spectra we compare to in this section are not
derived directly by the SSPP but, in most cases, they come from matching
templates from the Elodie library as part of the spectro-1d pipeline.
Nonetheless, the SSPP makes some choices regarding the adopted radial velocity,
as explained in Paper I and II, and therefore we refer to the finally adopted
radial velocity for the SDSS spectra as the SSPP values below.  

\begin{figure*}[ht!]
\centering
\includegraphics[width=14.cm,angle=0]{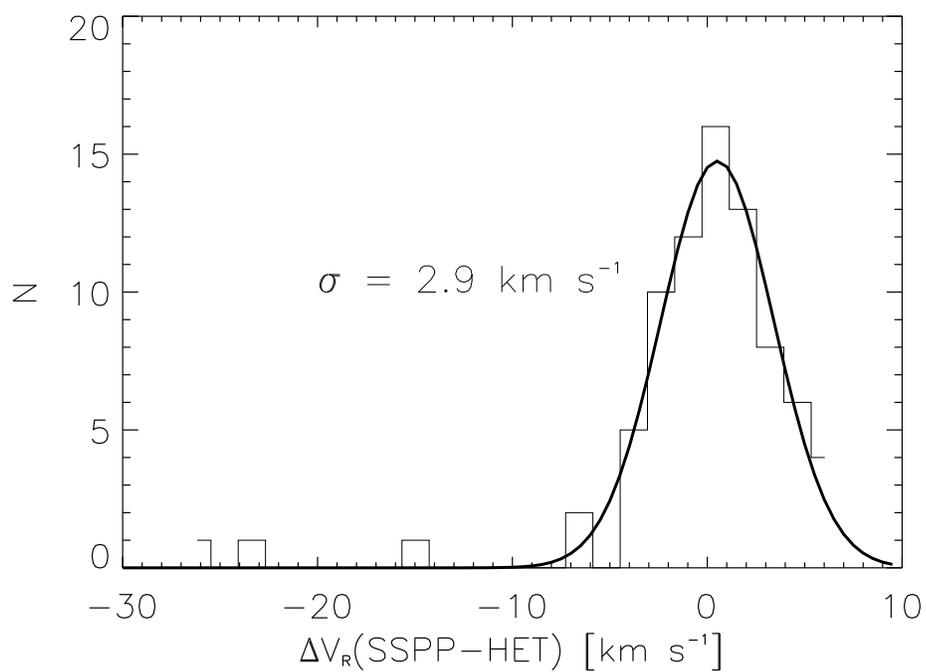}
\protect\caption[ ]{
Histogram of the differences between the radial velocities determined
by the SSPP from SDSS spectra, and those measured on the HET-HRS spectra.
The solid line illustrates a Gaussian model fit by least-squares to the data.
The three outliers are  SDSS  J233852.54+140945.7, SDSS  J013627.14+231453.6, 
and SDSS  J012106.89+263648.0.
}
\label{velo}
\end{figure*}

The mean
$S/N$ per pixel of the SDSS/SEGUE spectra in this set is
typically higher than 50/1.
The SSPP radial velocities exhibit an rms scatter of 5.1 km s$^{-1}$ 
relative to the average of our two preferred methods. Nevertheless, 
this value is not representative for most stars, but
it is inflated by three outliers 
(SDSS  J233852.54+140945.7, SDSS  J013627.14+231453.6, 
and SDSS  J012106.89+263648.0).
A more reliable indication of the typical
scatter is derived by least-squares fitting of a Normal curve to the differences,
which, as Figure \ref{velo} illustrates, yields  $\sigma= 2.9$ km s$^{-1}$.
This indicates a typical uncertainty
of about 2.4 km s$^{-1}$ for the SSPP radial velocities. 
This level of accuracy is better than in
earlier public data releases because of improvements to the DR6
version of the spectro-1d pipeline, primarily to the wavelength
solutions, and is consistent with the estimated the plate-to-plate scatter
in the radial velocity zero point (Adelman-McCarthy et al. 2007).

The SDSS radial velocities involved in our comparison have been systematically
corrected by $+7.3$ km s$^{-1}$, based on preliminary results from this program,
as described by Adelman-McCarthy et al. (2007), 
and therefore we limit our discussion to the variance. 
The unusually large errors for a few stars are likely related to some
issue with the SSPP or the SDSS spectra rather than on the HET side.
There are a few more examples among the stars observed with KeckI-ESI.
The (internal) error bars delivered by the SSPP for the stars in the HET sample
range between 0.7 and 2.0 km s$^{-1}$, with a mean value of about 1.3 km
s$^{-1}$, or about half our empirical external estimate.

\subsection{Atmospheric parameters}

The SSPP parameters derived for SDSS spectra discussed in this section are the
average values provided as part of SDSS DR-6 in the public data base
(Adelman-McCarthy et al. 2007). In Paper I, we compare the high-resolution
parameters against the individual methods integrated into the SSPP, in order to
estimate their associated systematic and random errors. These will be used in
future updates of the SSPP to weight the results from individual methods
appropriately.

\begin{figure*}[ht!]
\centering
\includegraphics[width=8.cm,angle=0]{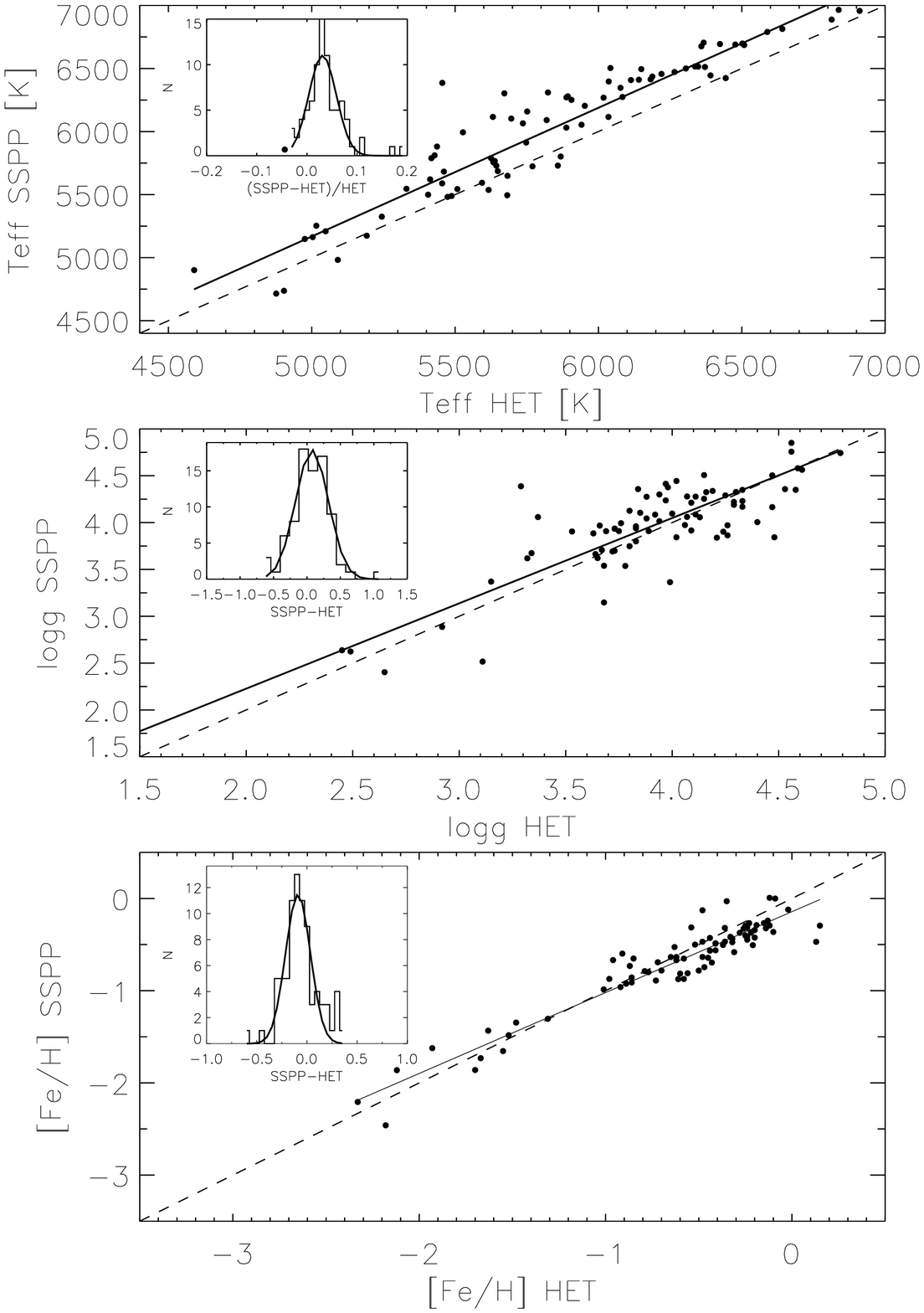}
\includegraphics[width=8.cm,angle=0]{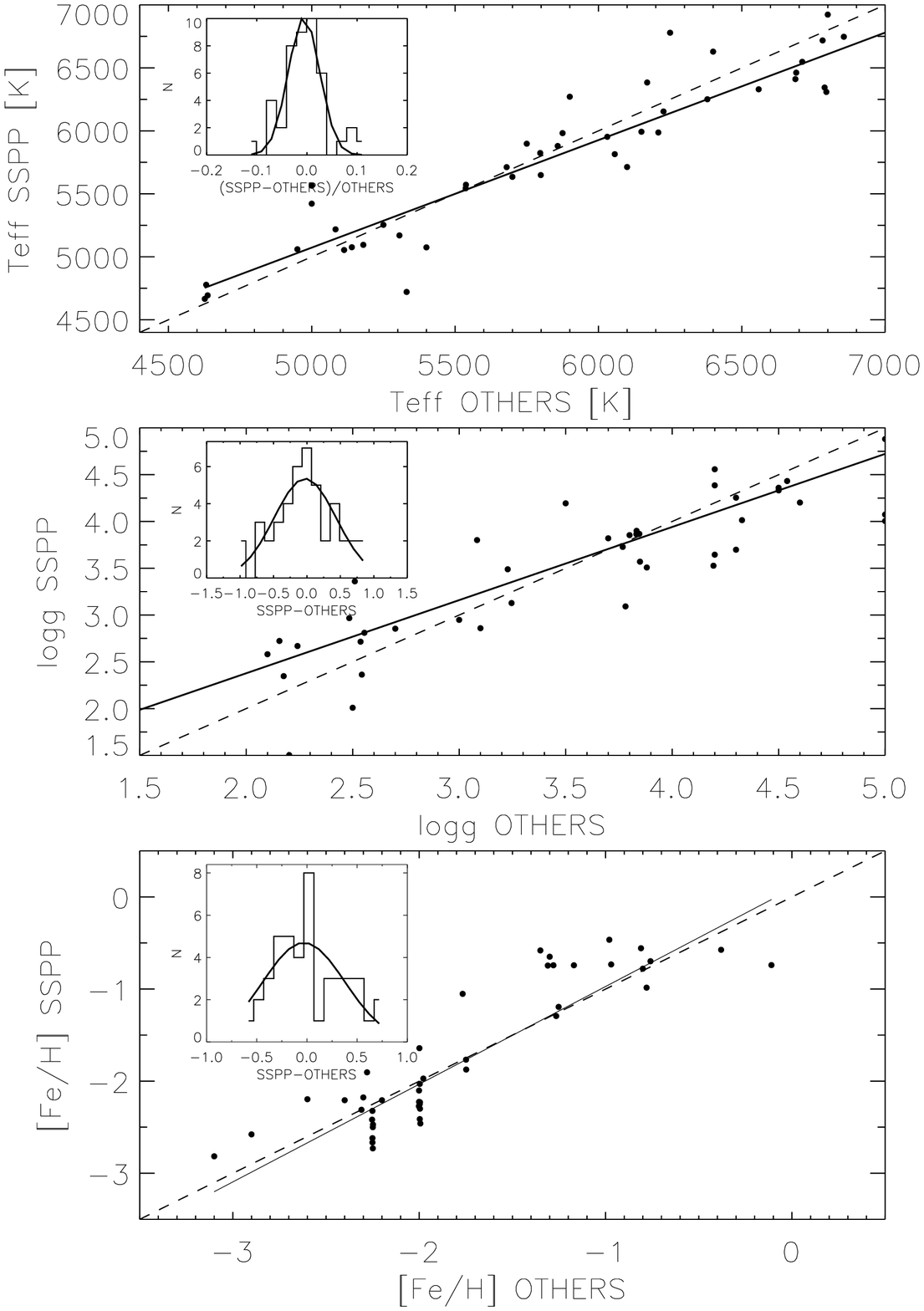}
\protect\caption[ ]{
Comparison between our derived parameters for the high-resoltuion 
spectra and the results of the SSPP based on SDSS data. The dashed lines indicate a slope of unity; the
solid lines are linear fits to the data. The insets show the
differences between the fit and catalog parameters, as well as Gaussian
models employed to make robust estimates of the median and standard
deviation, as shown in Table \ref{libraries}.   
}
\label{pipes}
\end{figure*}

Figure \ref{pipes} shows the main result of this paper, the comparison between
the estimated stellar atmospheric parameters obtained from the high-resolution
spectra with those from the SSPP based on SDSS data. Table \ref{compare}
summarizes the mean and standard deviation of the differences between the
high-resolution results (HI) and those from the SSPP. 

There is better agreement between the zero points of the SSPP parameters and the
high-resolution results for the OTHERS sample than for the HET results. However,
the rms scatter is significant smaller for the HET sample than for the OTHERS
sample. Despite the fact that we have chosen the high (spectroscopic) $T_{\rm
eff}$ scale for calibrating the HET results, we find that the SSPP indicates
even higher temperatures, by about 170~K; this value is comparable to the
scatter found for this parameter. For SDSS J180922.45+223712, the star observed
both with HET and Subaru, the HET and OTHERS analyses yielded disparate
effective temperatures of 5906 K and 6380 K, surface gravities of 4.40 dex and
5.00 dex, and metallicities of $-2.33$ and $-2.20$, respectively. The SSPP
$T_{\rm eff}$ estimate is intermediate to the two values, 6252 K. We note that
this is one of the stars with the lowest $S/N$ among the HET sample.

The larger scatter for the OTHERS sample is not attributable to the more
extended coverage of the parameter space; if we restrict the OTHERS sample to
the same range covered by the HET observations, the results do not vary
significantly. It is probably related to the different analysis techniques. For
example, the Keck-HIRES and Subaru analysis employs Fe I and Fe II lines, which
are mainly in the region around 390--450 nm, were the $S/N$ is lower than in the
redder region where the HET analysis is performed. The traditional analysis of
Fe I and Fe II lines uses weak lines, while the HET analysis also includes
stronger features, which may be more reliable at low $S/N$. In addition, the
effect of microturbulence is explicitly considered in the traditional analysis,
while the HET (and also ESI), as well as the SSPP techniques, consider a fixed
microturbulence value, and therefore are likely to be on the same scale. In
addition, the residuals for the OTHERS sample appear markedly non-Gaussian, and
the scatter determined from Gaussian fittings enhances the estimated width of
the distributions. In particular, the $\sigma_{\rm rms}$ for $\log g$ is 0.35
dex, while the value estimated from a Gaussian model is 0.41 dex (Table
\ref{compare}).

As previously explained, the uncertainties for the HET spectra are determined by
adding in quadrature the uncertainties inferred from the comparison with the
S$^4$N library, as shown in Table \ref{libraries}, and the $1\sigma$ scatter
among the values derived from the analysis of individual exposures of each star.
The latter contribution is, for most stars, negligible. The SSPP uncertainties
correspond to the standard error of the mean ($\sigma/\sqrt{N}$) for the results
from the different methods assembled in the pipeline. 

\begin{figure}[ht!]
\centering
\includegraphics[width=12.cm,angle=0]{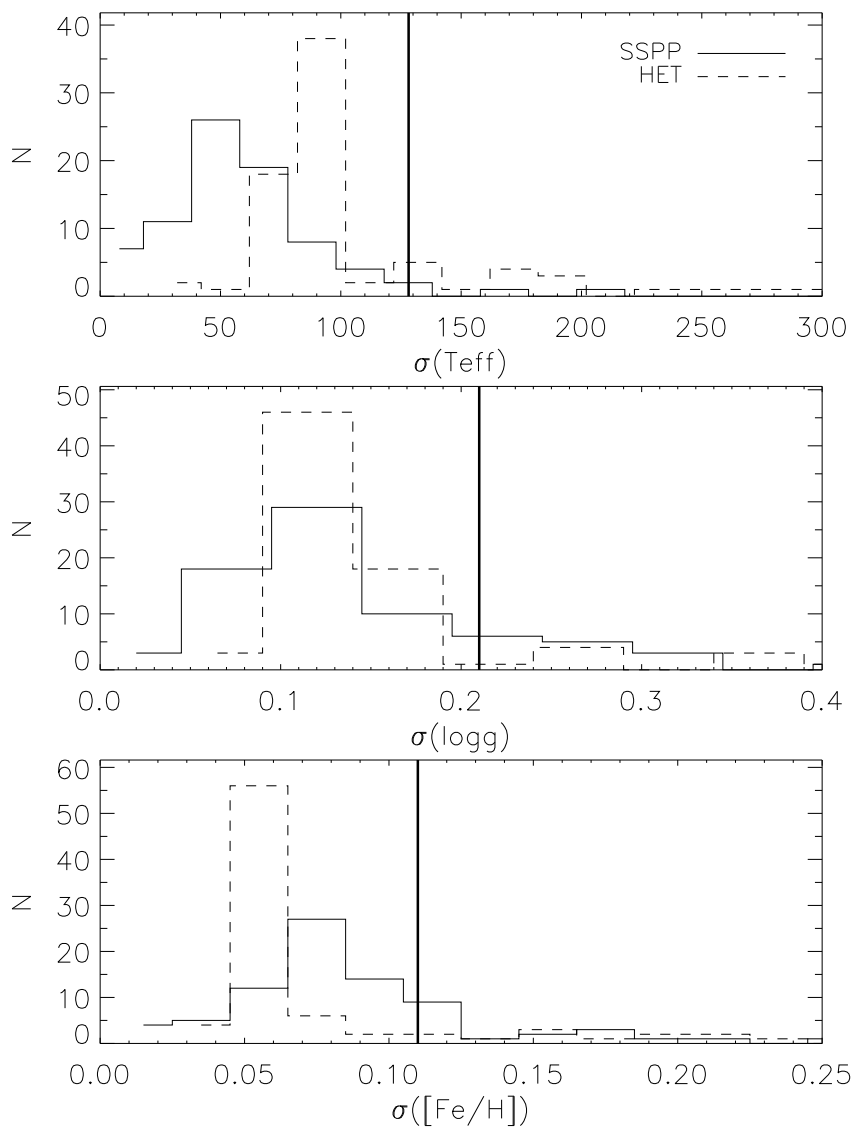}
\protect\caption[ ]{
Distribution of estimated (internal) uncertainties in the SSPP parameters (solid
lines) and
those from high-resolution spectroscopy (dashed) for the HET sample.
The vertical lines mark the more realistic external errors
for the SSPP parameters, as empirically derived from the comparison with our
analysis of HET spectra. }
\label{error}
\end{figure}

Figure \ref{error} shows histograms of the distribution of uncertainty estimates
in the HET sample for both the SSPP (solid lines) and the high-resolution HET
data (dashed lines). It is unlikely that the parameters obtained from our
analysis of high-resolution spectra are more uncertain than those reported by
the SSPP. The vertical lines indicate the empirical estimates derived for the
SSPP parameters from the comparison with the HET values (see Table
\ref{compare}).  The conclusion from this comparison is that the (internal)
SSPP error bars significantly underestimate the actual uncertainties, at least
for the SDSS/SEGUE spectra with relatively high signal-to-noise ratio ($ >
50/1$). Typically, the quoted SSPP uncertainties in the effective temperature
($\sim 50$ K) are about $2-3$ times too small, while those in surface gravity
($\sim 0.12$ dex) and metallicity ($\sim 0.08$ dex) are about half of their
actual external errors.

\section{Uncertainties as a function of S/N}

The comparison in Figure \ref{pipes} and Table \ref{compare} involves a
set of SDSS/SEGUE spectra with quite high $S/N$. Nevertheless, most of
the stellar spectra acquired in these projects have a significantly lower $S/N$
ratio, typically with a wavelength-averaged value of $10/1--30/1$. To estimate
the effect of lower $S/N$ on the derived atmospheric parameters, we have
introduced noise into the original observations. 

We followed the same recipe described by Allende Prieto (2007) to create new
sets of spectra degraded to a $S/N$ at 500 nm ($S/N_{\rm 500}$) of 5/1, 10/1,
20/1, and 40/1. All sets were analyzed using only one of the methods included in
the SSPP: spectral fitting with the {\tt ki13} grid, which is described in
Paper I (see also Allende Prieto et al. 2006). We found that the derived
effective temperatures agree with those determined from HET spectra with an rms
scatter of 13\%, 5\%, 4 \%, and 3.2 \% at $S/N_{\rm 500}$ of 5/1, 10/1, 20/1,
and 40/1, respectively. The derived surface gravities agreed with the
high-resolution values with an rms scatter of 0.70, 0.55, 0.42, and 0.30 dex,
while the metallicities agreed with an rms scatter of 0.71 dex, 0.29 dex, 0.15
dex, and 0.13 dex for a $S/N_{\rm 500}$ of 5/1, 10/1, 20/1, and 40/1,
respectively. Because the {\tt ki13} method and the HET analysis share a number
of elements (search algorithm and code, and spectral synthesis data and code),
and the spectral windows they exploit overlap, uncertainties could be slightly
underestimated at high $S/N$, but the figures derived at $S/N=40/1$ are in line
with those for the original SDSS spectra analyzed with the SSPP (Table
\ref{compare}).
 
\section{Conclusions}

We report on an analysis of high-resolution spectroscopic observations of a
sample of stars previously observed with the SDSS instrumentation as part of
SDSS-I or SEGUE. These new data are used to derive radial velocities and
atmospheric parameters, and to scrutinize the performance of the SSPP Pipeline
described in Paper I in this series. The sample we have examined includes 81
stars observed with the HET-HRS, 25 stars obtained with Keck-ESI, 11 stars
observed with Keck-HIRES, and 9 stars from Subaru-HDS.

Through a comparison with external spectroscopic libraries, and by employing
multiple methods of analysis for the HET sample, we estimate that our reference
radial velocities are accurate to 1.6 km s$^{-1}$. Our values for the stellar
atmospheric parameters, effective temperature, surface gravity, and metallicity,
are accurate to 1.5 \% ($\sim 90$~K), 0.13~dex and 0.05~dex, respectively. These
figures are derived from the comparison with the parameters for nearby stars in
the S$^4$N catalog, but we find they are still valid for the moderately high
$S/N$ of the HET spectra. Using the HET sample to benchmark the SSPP,
subtracting in quadrature the uncertainties in the results for the former, we
conclude that the SDSS/SEGUE radial velocities are typically accurate to 2.4 km
s$^{-1}$ for high signal-to-noise SDSS spectra ($S/N > 50/1$). A similar comparison
of the atmospheric parameters returned by the SSPP with those obtained from HET
spectra leads to the conclusion that the SSPP effective temperatures, surface
gravities, and metallicities for bright targets show random errors of 2.2\%
($\sim 130$ K), 0.21 dex, and 0.11 dex, respectively. Systematic offsets of a
similar size are detected for the effective temperatures and metallicities. We
evaluate the expected random uncertainties as a function of $S/N$ by repeating
the analysis after introducing noise in the SDSS spectra. More extended tests
are underway and will be reported elsewhere.

Our study also finds that the internal uncertainties delivered by the SSPP
for both radial velocities and atmospheric parameters need to be systematically
increased by a factor of $2-3$ in order to be consistent with our derived external
errors. The uncertainties in the average SSPP atmospheric parameters are simply
derived as the standard error of the mean for a Normal distribution from the
multiple techniques applied to any particular target. The fact that many methods
share the same spectroscopic indicators (e.g. Balmer lines or SDSS color indices
to gauge $T_{\rm eff}$), and models (e.g. Kurucz's model atmospheres) may cause
unaccounted correlations that result in underestimated uncertainties.  

The validation and calibration of the SSPP is an ongoing project. Several
additional open and globular clusters have recently had data obtained with SDSS
instrumentation, and will be considered in future papers. A sample of up to
several hundred very low-metallicity stars from SDSS/SEGUE is presently being
observed with the HET, which we will add to our calibration sample. Additional
stars of intermediate metallicity, and with hotter and cooler temperatures than
considered in the present work, will be added to our calibration
sample based on observations with a number of large-aperture telescopes.
Our goal is to produce an SSPP validation catalog for on
the order of 500 stars, which will be used to refine and adjust the individual
parameter estimation techniques employed by the SSPP, and thus establish a
definitive atmospheric parameter estimation scale for application to the large
(and growing) SDSS/SEGUE stellar samples, as well as to other future surveys. 

\acknowledgments

Funding for the SDSS and SDSS-II has been provided by the Alfred P. Sloan
Foundation, the Participating Institutions, the National Science Foundation, the
U.S. Department of Energy, the National Aeronautics and Space Administration,
the Japanese Monbukagakusho, the Max Planck Society, and the Higher Education
Funding Council for England. The SDSS Web Site is http://www.sdss.org/.

The SDSS is managed by the Astrophysical Research Consortium for the
Participating Institutions. The Participating Institutions are the American
Museum of Natural History, Astrophysical Institute Potsdam, University of Basel,
University of Cambridge, Case Western Reserve University, University of Chicago,
Drexel University, Fermilab, the Institute for Advanced Study, the Japan
Participation Group, Johns Hopkins University, the Joint Institute for Nuclear
Astrophysics, the Kavli Institute for Particle Astrophysics and Cosmology, the
Korean Scientist Group, the Chinese Academy of Sciences (LAMOST), Los Alamos
National Laboratory, the Max-Planck-Institute for Astronomy (MPIA), the
Max-Planck-Institute for Astrophysics (MPA), New Mexico State University, Ohio
State University, University of Pittsburgh, University of Portsmouth, Princeton
University, the United States Naval Observatory, and the University of
Washington.

The Hobby-Eberly Telescope (HET) is a joint project of the University of Texas
at Austin, the Pennsylvania State University, Stanford University,
Ludwig-Maximilians-Universit\"at M\"unchen, and Georg-August-Universit\"at
G\"ottingen. The HET is named in honor of its principal benefactors, William P.
Hobby and Robert E. Eberly. Some of the data presented herein were obtained at
the W.M. Keck Observatory, which is operated as a scientific partnership among
the California Institute of Technology, the University of California and the
National Aeronautics and Space Administration. The Observatory was made possible
by the generous financial support of the W.M. Keck Foundation. The authors wish
to recognize and acknowledge the very significant cultural role and reverence
that the summit of Mauna Kea has always had within the indigenous Hawaiian
community. We are most fortunate to have the opportunity to conduct observations
from this mountain.

NASA grants (NAG5-13057, NAG5-13147) to C.~A.~P.  and D. L. L.
 are thankfully acknowledged.
T.~C.~B., Y.~S.~L., B.~M., and T.~S. acknowledge support from the US National
Science Foundation under grants AST 04-06784 and AST 07-07776, as well as from
grant PHY 02-16783; Physics Frontier Center/Joint Institute for Nuclear
Astrophysics (JINA). D. L. L.'s research is supported in part by the
Welch Foundation of Houston, Texas.


\clearpage

\clearpage
\begin{landscape}

\begin{deluxetable}{lrrclll}
\tablecolumns{7}
\tablewidth{0pc} 
\tablecaption{Observations}
\tablehead{
\colhead{Telescope}        & \colhead{Instrument}    &   \colhead{Resolving Power}  & 
\colhead{Slit Width [arcsec]}  & 
\colhead{Wavelength Coverage [nm]}   & \colhead{S/N [per pixel]} & \colhead{No. of objects}   }
\startdata
  HET              & HRS        &   15000       & 2.00 & 450-770   & 20-50   & 81 \\
  Keck I           & HIRES      &   40000       & 0.86 & 414-850   & 80      & 11 \\
  Keck II          & ESI        &    7000       & 0.75 & 380-1000  & 30-60   & 25 \\
  Subaru           & HDS        &   50000       & 0.72 & 300-580   & 80      & 9  	
\enddata
\label{table1}
\end{deluxetable}

\end{landscape}

\begin{deluxetable}{rrrr}
\tablecolumns{4}
\tablewidth{0pc} 
\tablecaption{Average and $1\sigma$ Scatter Between
Derived Parameters (FIT) and the Library Catalogs (LIB)}
\tablehead{
\colhead{Parameter/Library} & \colhead{$<{\rm FIT} - {\rm LIB}>$}	& 
\colhead{$\sigma({\rm FIT} - {\rm LIB})$}  & \colhead{N} }
\startdata
	$T_{\rm eff}$-S$^4$N (K)           &   $-0.10$\% &    1.67\%	&	55 \\
	
	log $g$-S$^4$N (dex)	  &	0.008	&    0.129	&	   \\
	
	[Fe/H]-S$^4$N  (dex)	  &    $-0.001$  &    0.049	&	   \\
	
\tableline

	$T_{\rm eff}$-Elodie (K)  	  &   $-2.23$\%	&    2.66\%	&      282 \\
	
	log $g$-Elodie (dex)	  &	0.017	&    0.271	&	   \\
	
	[Fe/H]-Elodie  (dex)	  &   $-0.020$	&    0.100	&	   \\

\enddata
\label{libraries}
\end{deluxetable}

\begin{landscape}


\begin{deluxetable}{rrrrrrrrrrrrrrrrr} 
\tabletypesize{\scriptsize}
\tablecolumns{17}
\tablewidth{0pc} 
\tablecaption{Comparison of SSPP Velocities and Atmospheric Parameters (HET
Sample)} 
\tablehead{ 
\colhead{} & \colhead{}   &  \multicolumn{7}{c}{SSPP} &   \colhead{}   & 
\multicolumn{7}{c}{HET} \\ 
\cline{3-9} \cline{11-17} \\ 
\colhead{Star} & \colhead{MJD-PLATE-FIBER}  & \colhead{$V_R$}   
& \colhead{$T_{\rm eff}$}    & \colhead{$\sigma$}    
& \colhead{$\log g$} 	     & \colhead{$\sigma$} 
& \colhead{[Fe/H]}  & \colhead{$\sigma$}      
& \colhead{} & \colhead{$V_R$}   
& \colhead{$T_{\rm eff}$}    & \colhead{$\sigma$}    
& \colhead{$\log g$} 	     & \colhead{$\sigma$} 
& \colhead{[Fe/H]}  & \colhead{$\sigma$}      }
\startdata 
SDSS  J171652.50+603926.9	 & 	51703-0353-605	 & $-$53.60 & 6303 & 158 & 4.06 & 0.03 & $-$0.03 & 0.05 & & $-$50.54 & 5672 & 82 & 3.37 & 0.13 & $-$0.35 & 0.05\\
SDSS  J225801.77+000643.1	 & 	51792-0380-236	 & 3.80 & 6965 & 88 & 3.97 & 0.31 & $-$0.58 & 0.01 & & 3.97 & 6838 & 99 & 4.26 & 0.13 & $-$0.31 & 0.05\\
SDSS  J010746.51+011402.6	 & 	51816-0396-605	 & $-$40.60 & 5682 & 475 & 4.74 & 0.12 & 0.01 & 0.11 & & $-$43.37 & 5461 & 180 & 4.79 & 0.13 & $-$0.12 & 0.07\\
SDSS  J014149.73+010720.2	 & 	51788-0401-407	 & 15.70 & 4715 & 85 & 3.37 & 0.70 & $-$0.64 & 0.01 & & 15.81 & 4876 & 73 & 3.15 & 0.16 & $-$0.45 & 0.06\\
SDSS  J014215.40+011400.6	 & 	51788-0401-410	 & 43.50 & 5789 & 79 & 4.38 & 0.12 & $-$0.31 & 0.10 & & 43.68 & 5417 & 82 & 3.98 & 0.13 & $-$0.54 & 0.06\\
SDSS  J024740.30+011144.9	 & 	51871-0409-449	 & $-$8.90 & 5802 & 60 & 4.56 & 0.11 & $-$0.12 & 0.09 & & $-$10.17 & 5868 & 87 & 4.61 & 0.13 & $-$0.02 & 0.05\\
SDSS  J025046.89+010910.8	 & 	51871-0409-562	 & $-$70.20 & 5994 & 61 & 4.36 & 0.12 & $-$0.73 & 0.13 & & $-$76.74 & 5527 & 187 & 3.84 & 0.28 & $-$0.87 & 0.16\\
SDSS  J005826.06+150153.6	 & 	51821-0421-439	 & $-$28.10 & 5163 & 28 & 3.62 & 0.73 & $-$0.32 & 0.07 & & $-$28.51 & 5003 & 162 & 3.32 & 0.45 & $-$0.24 & 0.09\\
SDSS  J074705.19+414452.1	 & 	51885-0434-133	 & 73.90 & 5209 & 206 & 3.67 & 0.42 & $-$0.42 & 0.11 & & 77.84 & 5048 & 74 & 3.34 & 0.13 & $-$0.33 & 0.05\\
SDSS  J082253.87+471742.0	 & 	51868-0441-497	 & $-$8.00 & 6504 & 111 & 3.36 & 0.63 & $-$0.64 & 0.09 & & $-$5.67 & 6042 & 87 & 3.99 & 0.13 & $-$0.65 & 0.05\\
SDSS  J103146.22+012710.5	 & 	52316-0504-016	 & 12.30 & 5881 & 130 & 3.99 & 0.01 & $-$0.67 & 0.17 & & 14.64 & 5437 & 45 & 3.76 & 0.07 & $-$0.96 & 0.04\\
SDSS  J115520.82+654309.8	 & 	52316-0598-443	 & $-$10.40 & 6116 & 74 & 4.13 & 0.11 & $-$0.92 & 0.20 & & $-$7.01 & 5632 & 85 & 3.80 & 0.13 & $-$0.89 & 0.05\\
SDSS  J134901.58+640924.7	 & 	52079-0604-572	 & $-$3.00 & 6091 & 5 & 4.33 & 0.30 & $-$0.00 & 0.01 & & 3.91 & 5819 & 92 & 4.16 & 0.15 & $-$0.09 & 0.06\\
SDSS  J151241.70+593151.5	 & 	52345-0613-280	 & $-$58.00 & 5649 & 117 & 4.36 & 0.23 & $-$0.60 & 0.08 & & $-$61.13 & 5683 & 89 & 4.53 & 0.13 & $-$0.91 & 0.05\\
SDSS  J213818.93+123547.8	 & 	52221-0732-345	 & $-$45.70 & 5174 & 44 & 3.70 & 0.11 & $-$0.78 & 0.01 & & $-$44.19 & 5192 & 77 & 3.73 & 0.13 & $-$0.70 & 0.05\\
SDSS  J224610.22+145156.7	 & 	52263-0740-364	 & 23.20 & 6386 & 674 & 4.09 & 0.60 & $-$0.13 & 0.08 & & 19.17 & 5455 & 325 & 4.11 & 0.57 & $-$0.48 & 0.23\\
SDSS  J231427.17+134821.9	 & 	52251-0744-179	 & $-$82.30 & 4982 & 71 & 3.15 & 0.73 & $-$0.37 & 0.08 & & $-$82.23 & 5091 & 74 & 3.68 & 0.14 & $-$0.28 & 0.05\\
SDSS  J233720.38+140953.8	 & 	52234-0747-136	 & $-$12.70 & 6424 & 39 & 4.17 & 0.16 & $-$0.32 & 0.07 & & $-$12.15 & 6444 & 120 & 4.47 & 0.18 & $-$0.14 & 0.05\\
SDSS  J233611.87+140923.9	 & 	52234-0747-212	 & $-$5.40 & 6054 & 2 & 3.97 & 0.12 & $-$0.69 & 0.06 & & $-$5.64 & 5941 & 86 & 4.06 & 0.13 & $-$0.43 & 0.05\\
SDSS  J124826.99+614358.8	 & 	52373-0781-015	 & $-$26.70 & 6204 & 38 & 4.24 & 0.11 & $-$0.44 & 0.10 & & $-$20.32 & 5952 & 86 & 3.97 & 0.13 & $-$0.32 & 0.05\\
SDSS  J155509.18+495003.3	 & 	52352-0812-578	 & $-$50.60 & 5766 & 34 & 4.35 & 0.14 & $-$0.50 & 0.07 & & $-$48.44 & 5638 & 82 & 4.33 & 0.13 & $-$0.37 & 0.05\\
SDSS  J111901.08+054319.4	 & 	52326-0835-601	 & $-$2.60 & 5729 & 67 & 4.50 & 0.22 & $-$0.37 & 0.08 & & $-$4.64 & 5644 & 81 & 4.47 & 0.13 & $-$0.22 & 0.05\\
SDSS  J074151.21+275319.8	 & 	52339-0888-599	 & 6.40 & 7171 & 52 & 3.84 & 0.12 & $-$0.42 & 0.05 & & 12.77 & 6475 & 93 & 4.48 & 0.13 & $-$0.20 & 0.05\\
SDSS  J074300.91+285106.6	 & 	52663-0889-204	 & 51.00 & 6515 & 129 & 4.29 & 0.11 & $-$0.78 & 0.17 & & 53.38 & 6348 & 92 & 4.25 & 0.13 & $-$0.50 & 0.05\\
SDSS  J112848.08+580740.4	 & 	52409-0952-260	 & $-$12.40 & 5811 & 63 & 4.44 & 0.20 & $-$0.47 & 0.08 & & $-$13.56 & 5428 & 283 & 4.02 & 0.47 & $-$0.48 & 0.21\\
SDSS  J161511.43+352900.2	 & 	52764-1056-124	 & 31.50 & 6457 & 38 & 4.10 & 0.13 & $-$0.24 & 0.07 & & 32.17 & 6220 & 90 & 4.00 & 0.13 & $-$0.13 & 0.05\\
SDSS  J235427.13+351233.4	 & 	53262-1880-087	 & $-$56.30 & 6271 & 45 & 3.71 & 0.25 & $-$1.66 & 0.04 & & $-$57.76 & 5888 & 99 & 3.67 & 0.14 & $-$1.55 & 0.05\\
SDSS  J234952.45+365447.3	 & 	53262-1880-428	 & $-$80.90 & 6269 & 51 & 3.84 & 0.09 & $-$1.73 & 0.05 & & $-$85.73 & 6018 & 87 & 4.21 & 0.13 & $-$1.67 & 0.05\\
SDSS  J233946.60+433049.4	 & 	53228-1884-428	 & 1.10 & 6690 & 62 & 4.25 & 0.09 & $-$0.30 & 0.07 & & 0.64 & 6477 & 98 & 4.15 & 0.13 & $-$0.25 & 0.05\\
SDSS  J211622.82+114002.5	 & 	53237-1890-527	 & 4.80 & 6031 & 61 & 4.33 & 0.14 & $-$0.27 & 0.09 & & 3.79 & 5888 & 101 & 4.30 & 0.14 & $-$0.23 & 0.05\\
SDSS  J233852.54+140945.7	 & 	53240-1894-296	 & $-$269.90 & 5253 & 50 & 2.62 & 0.17 & $-$1.48 & 0.10 & & $-$243.22 & 5016 & 72 & 2.49 & 0.13 & $-$1.52 & 0.05\\
SDSS  J004436.24+160203.6	 & 	53242-1896-445	 & $-$138.80 & 5621 & 32 & 3.94 & 0.18 & $-$0.65 & 0.06 & & $-$139.80 & 5413 & 78 & 3.83 & 0.13 & $-$0.85 & 0.05\\
SDSS  J004416.51+244246.6	 & 	53327-2038-154	 & $-$303.10 & 5594 & 56 & 2.52 & 0.33 & $-$2.46 & 0.04 & & $-$304.35 & 5594 & 197 & 3.11 & 0.14 & $-$2.18 & 0.18\\
SDSS  J003916.49+242339.5	 & 	53327-2038-226	 & $-$5.00 & 6512 & 38 & 4.21 & 0.08 & $-$0.47 & 0.05 & & $-$0.74 & 6371 & 95 & 4.09 & 0.13 & $-$0.36 & 0.05\\
SDSS  J003749.37+252708.4	 & 	53327-2038-382	 & $-$15.20 & 6411 & 27 & 4.08 & 0.08 & $-$0.64 & 0.04 & & $-$15.61 & 6141 & 89 & 3.92 & 0.13 & $-$0.62 & 0.05\\
SDSS  J004537.38+253506.3	 & 	53327-2038-564	 & 14.60 & 6159 & 3 & 3.62 & 0.21 & $-$0.69 & 0.08 & & 15.06 & 5751 & 175 & 3.65 & 0.34 & $-$0.72 & 0.14\\
SDSS  J011751.77+243604.4	 & 	53384-2040-083	 & $-$85.80 & 5857 & 67 & 4.39 & 0.22 & $-$1.62 & 0.10 & & $-$84.87 & 4906 & 737 & 3.29 & 1.07 & $-$1.93 & 0.38\\
SDSS  J012049.43+254940.8	 & 	53384-2040-595	 & $-$127.40 & 5482 & 50 & 2.89 & 0.15 & $-$0.53 & 0.07 & & $-$126.97 & 5474 & 79 & 2.92 & 0.13 & $-$0.63 & 0.05\\
SDSS  J012106.89+263648.0	 & 	53384-2040-617	 & $-$56.30 & 5730 & 41 & 4.22 & 0.07 & $-$0.29 & 0.30 & & $-$41.61 & 5858 & 87 & 4.29 & 0.13 & 0.15 & 0.05\\
SDSS  J012116.42+261354.0	 & 	53384-2040-637	 & $-$12.10 & 6495 & 35 & 4.04 & 0.04 & $-$0.67 & 0.04 & & $-$13.12 & 6149 & 129 & 3.88 & 0.18 & $-$0.62 & 0.08\\
SDSS  J012441.76+305553.3	 & 	53387-2041-008	 & 43.00 & 5544 & 22 & 3.66 & 0.07 & $-$0.81 & 0.08 & & 39.99 & 5508 & 79 & 3.64 & 0.13 & $-$0.60 & 0.05\\
SDSS  J012945.31+375221.6	 & 	53378-2042-009	 & 26.50 & 6500 & 52 & 4.06 & 0.12 & $-$0.74 & 0.12 & & 27.53 & 6305 & 91 & 4.07 & 0.13 & $-$0.47 & 0.05\\
SDSS  J012314.37+384749.1	 & 	53378-2042-461	 & $-$264.90 & 6445 & 93 & 3.80 & 0.26 & $-$1.86 & 0.11 & & $-$267.98 & 6391 & 92 & 3.83 & 0.13 & $-$1.70 & 0.05\\
SDSS  J013930.32+222533.4	 & 	53327-2044-122	 & 77.70 & 5546 & 49 & 4.51 & 0.11 & $-$0.96 & 0.15 & & 74.52 & 5330 & 79 & 4.15 & 0.14 & $-$0.92 & 0.05\\
SDSS  J013924.06+231006.8	 & 	53327-2044-167	 & $-$17.80 & 5686 & 29 & 4.19 & 0.12 & $-$0.34 & 0.10 & & $-$15.74 & 5649 & 99 & 4.29 & 0.15 & $-$0.20 & 0.06\\
SDSS  J013627.14+231453.6	 & 	53327-2044-228	 & $-$145.40 & 4901 & 92 & 1.60 & 0.14 & $-$1.86 & 0.26 & & $-$122.72 & 4590 & 66 & 1.49 & 0.13 & $-$2.12 & 0.05\\
SDSS  J021317.01+220622.7	 & 	53327-2046-061	 & 5.00 & 6516 & 37 & 4.28 & 0.08 & $-$0.56 & 0.06 & & 5.14 & 6337 & 70 & 4.11 & 0.08 & $-$0.41 & 0.04\\
SDSS  J033530.56$-$010038.3	 & 	53350-2049-020	 & 1.70 & 5756 & 10 & 4.23 & 0.13 & $-$0.81 & 0.09 & & $-$2.05 & 5633 & 150 & 4.33 & 0.28 & $-$0.56 & 0.07\\
SDSS  J032930.11$-$010721.1	 & 	53350-2049-241	 & 50.60 & 5589 & 52 & 4.11 & 0.11 & $-$0.45 & 0.16 & & 49.16 & 5455 & 37 & 3.85 & 0.06 & $-$0.24 & 0.03\\
SDSS  J053442.39+003826.7	 & 	53401-2052-533	 & 33.80 & 6116 & 112 & 4.35 & 0.19 & $-$0.43 & 0.12 & & 36.93 & 6035 & 27 & 4.58 & 0.15 & $-$0.44 & 0.04\\
SDSS  J073240.79+351717.7	 & 	53446-2053-023	 & 20.90 & 6676 & 47 & 3.91 & 0.13 & $-$0.50 & 0.05 & & 20.14 & 6359 & 338 & 3.75 & 0.56 & $-$0.52 & 0.19\\
SDSS  J073034.52+352545.9	 & 	53446-2053-130	 & 18.00 & 6279 & 65 & 4.28 & 0.11 & $-$0.80 & 0.09 & & 16.21 & 5893 & 138 & 3.88 & 0.14 & $-$0.77 & 0.11\\
SDSS  J072801.58+354503.3	 & 	53446-2053-171	 & 9.10 & 6347 & 45 & 3.88 & 0.08 & $-$0.89 & 0.07 & & 9.28 & 6077 & 88 & 3.63 & 0.13 & $-$0.73 & 0.05\\
SDSS  J072753.81+345437.5	 & 	53446-2053-226	 & $-$26.40 & 6790 & 84 & 3.92 & 0.26 & $-$0.57 & 0.08 & & $-$25.61 & 6589 & 99 & 4.09 & 0.13 & $-$0.44 & 0.07\\
SDSS  J072653.66+370019.9	 & 	53446-2053-346	 & 30.70 & 6814 & 72 & 4.06 & 0.09 & $-$0.49 & 0.04 & & 33.23 & 6641 & 132 & 4.13 & 0.13 & $-$0.41 & 0.07\\
SDSS  J072940.24+370322.8	 & 	53446-2053-505	 & $-$8.30 & 6686 & 66 & 3.86 & 0.28 & $-$0.47 & 0.07 & & $-$8.11 & 6508 & 94 & 4.26 & 0.14 & $-$0.32 & 0.05\\
SDSS  J074512.81+170144.2	 & 	53431-2054-056	 & 71.40 & 6957 & 50 & 3.90 & 0.09 & $-$0.27 & 0.07 & & 67.66 & 6911 & 100 & 4.24 & 0.13 & $-$0.15 & 0.05\\
SDSS  J074139.58+172517.2	 & 	53431-2054-259	 & 29.40 & 6888 & 39 & 3.75 & 0.06 & $-$0.42 & 0.06 & & 32.07 & 6813 & 122 & 3.80 & 0.34 & $-$0.24 & 0.07\\
SDSS  J074638.40+183420.8	 & 	53431-2054-552	 & 25.50 & 6414 & 34 & 3.94 & 0.07 & $-$0.91 & 0.06 & & 23.75 & 6182 & 89 & 3.73 & 0.13 & $-$0.86 & 0.05\\
SDSS  J074112.78+205959.2	 & 	53378-2078-014	 & 38.50 & 5724 & 61 & 3.54 & 0.15 & $-$0.87 & 0.07 & & 33.57 & 5770 & 85 & 3.78 & 0.13 & $-$0.61 & 0.05\\
SDSS  J074125.25+212940.8	 & 	53378-2078-040	 & 30.20 & 6694 & 48 & 3.91 & 0.13 & $-$0.33 & 0.07 & & 27.83 & 6424 & 93 & 3.69 & 0.13 & $-$0.26 & 0.05\\
SDSS  J074017.97+205439.5	 & 	53378-2078-044	 & 11.00 & 6698 & 35 & 3.96 & 0.10 & $-$0.29 & 0.07 & & 7.55 & 6502 & 94 & 3.83 & 0.13 & $-$0.19 & 0.05\\
SDSS  J073938.60+202314.9	 & 	53378-2078-049	 & 42.00 & 6436 & 92 & 4.01 & 0.16 & $-$0.65 & 0.07 & & 42.37 & 6188 & 89 & 3.94 & 0.13 & $-$0.58 & 0.05\\
SDSS  J073752.70+205855.3	 & 	53378-2078-136	 & 14.20 & 6397 & 45 & 3.90 & 0.08 & $-$0.86 & 0.06 & & 12.70 & 6036 & 87 & 3.53 & 0.13 & $-$0.86 & 0.05\\
SDSS  J074010.36+213755.0	 & 	53378-2078-598	 & 32.50 & 6408 & 12 & 3.97 & 0.10 & $-$0.79 & 0.05 & & 33.91 & 6113 & 190 & 3.66 & 0.25 & $-$0.79 & 0.16\\
SDSS  J074125.25+212940.8	 & 	53379-2079-040	 & 31.60 & 6705 & 52 & 3.84 & 0.08 & $-$0.27 & 0.08 & & 28.54 & 6367 & 92 & 4.02 & 0.13 & $-$0.23 & 0.05\\
SDSS  J165640.62+393244.5	 & 	53524-2181-218	 & 16.90 & 5537 & 38 & 4.58 & 0.09 & $-$0.36 & 0.12 & & 12.44 & 5617 & 82 & 4.59 & 0.13 & $-$0.10 & 0.05\\
SDSS  J174638.20+243308.0	 & 	53536-2183-131	 & 5.40 & 5495 & 46 & 4.17 & 0.09 & $-$0.47 & 0.21 & & 7.11 & 5682 & 177 & 4.33 & 0.26 & 0.13 & 0.12\\
SDSS  J174431.30+252145.3	 & 	53536-2183-197	 & $-$52.40 & 5913 & 7 & 3.69 & 0.12 & $-$0.87 & 0.10 & & $-$51.91 & 5748 & 85 & 3.72 & 0.14 & $-$0.98 & 0.05\\
SDSS  J180922.45+223712.4	 & 	53534-2184-058	 & $-$366.10 & 6251 & 75 & 4.00 & 0.27 & $-$2.21 & 0.11 & & $-$371.02 & 5906 & 226 & 4.40 & 0.19 & $-$2.33 & 0.15\\
SDSS  J180831.36+223720.1	 & 	53534-2184-083	 & $-$71.60 & 6065 & 74 & 3.93 & 0.20 & $-$0.32 & 0.08 & & $-$77.42 & 5736 & 259 & 3.73 & 0.37 & $-$0.36 & 0.19\\
SDSS  J180924.48+231156.0	 & 	53534-2184-107	 & $-$200.10 & 5148 & 42 & 2.64 & 0.19 & $-$1.35 & 0.10 & & $-$195.85 & 4976 & 72 & 2.45 & 0.13 & $-$1.48 & 0.05\\
SDSS  J180534.75+244052.7	 & 	53534-2184-413	 & $-$45.10 & 5489 & 69 & 4.76 & 0.10 & $-$0.29 & 0.10 & & $-$44.64 & 5488 & 80 & 4.56 & 0.14 & $-$0.12 & 0.05\\
SDSS  J180418.33+234842.1	 & 	53534-2184-429	 & $-$166.20 & 6310 & 75 & 4.41 & 0.18 & $-$1.43 & 0.07 & & $-$170.78 & 5824 & 84 & 3.97 & 0.13 & $-$1.63 & 0.05\\
SDSS  J180623.33+245131.0	 & 	53534-2184-451	 & 9.20 & 6274 & 60 & 4.34 & 0.12 & $-$0.49 & 0.07 & & 3.60 & 6084 & 103 & 4.19 & 0.13 & $-$0.41 & 0.09\\
SDSS  J202718.90+125957.9	 & 	53558-2248-060	 & 54.20 & 5789 & 22 & 4.28 & 0.08 & $-$0.63 & 0.10 & & 52.61 & 5625 & 277 & 4.07 & 0.43 & $-$0.48 & 0.22\\
SDSS  J202244.17+131606.3	 & 	53558-2248-221	 & $-$41.70 & 6103 & 65 & 4.30 & 0.14 & $-$1.30 & 0.12 & & $-$47.53 & 5696 & 75 & 3.94 & 0.09 & $-$1.31 & 0.05\\
SDSS  J202301.63+123634.9	 & 	53558-2248-247	 & 44.20 & 6472 & 86 & 3.91 & 0.10 & $-$0.40 & 0.08 & & 42.07 & 6265 & 131 & 3.89 & 0.18 & $-$0.25 & 0.06\\
SDSS  J202039.15+140755.2	 & 	53558-2248-345	 & $-$30.00 & 5499 & 106 & 4.85 & 0.08 & $-$0.51 & 0.11 & & $-$30.08 & 5406 & 78 & 4.56 & 0.13 & $-$0.21 & 0.05\\
SDSS  J220537.22+202904.8	 & 	53557-2251-305	 & 41.70 & 5325 & 38 & 3.54 & 0.18 & $-$0.99 & 0.07 & & 39.62 & 5244 & 77 & 3.68 & 0.14 & $-$1.01 & 0.05\\
SDSS  J012811.36+385641.0	 & 	53712-2336-052	 & $-$63.70 & 4737 & 4 & 2.40 & 0.48 & $-$0.87 & 0.17 & & $-$61.84 & 4903 & 72 & 2.65 & 0.14 & $-$0.58 & 0.05\\
\enddata
\label{params}
\end{deluxetable}

\begin{deluxetable}{rrrrrrrrrrrrrrrrr} 
\tabletypesize{\scriptsize}
\tablecolumns{17}
\tablewidth{0pc} 
\tablecaption{Comparison of SSPP Velocities and Atmospheric Parameters (OTHERS
Sample)} 
\tablehead{ 
\colhead{}    &  \colhead{}    &  \multicolumn{7}{c}{SSPP} &   \colhead{}   & 
\multicolumn{7}{c}{OTHERS} \\ 
\cline{3-9} \cline{11-17} \\
\colhead{Star} & \colhead{MJD-PLATE-FIBER} & \colhead{$V_R$}   
& \colhead{$T_{\rm eff}$}    & \colhead{$\sigma$}    
& \colhead{$\log g$} 	     & \colhead{$\sigma$} 
& \colhead{[Fe/H]}  & \colhead{$\sigma$}      
& \colhead{} & \colhead{$V_R$}   
& \colhead{$T_{\rm eff}$}    & \colhead{$\sigma$}    
& \colhead{$\log g$} 	     & \colhead{$\sigma$} 
& \colhead{[Fe/H]}  & \colhead{$\sigma$}      }
\startdata 
\multicolumn{17}{c}{Keck-HIRES}  \\
\tableline
SDSS  J131137.14+000803.4	 & 	51986-0294-623	 & $-$22.40 & 5060 & 75 & 2.95 & 0.12 & $-$0.58 & 0.02 & & $-$18.40 & 4950 & 190 & 3.00 & 0.15 & $-$1.35 & 0.04\\
SDSS  J132847.82+010708.6	 & 	51959-0297-569	 & $-$48.60 & 5565 & 44 & 2.85 & 0.27 & $-$1.90 & 0.06 & & $-$45.60 & 5000 & 70 & 2.70 & 0.06 & $-$2.28 & 0.01\\
SDSS  J135432.19+000511.3	 & 	51943-0300-038	 & 34.80 & 5898 & 43 & 4.39 & 0.15 & $-$0.75 & 0.10 & & 34.50 & 5750 & 100 & 4.20 & 0.08 & $-$1.31 & 0.02\\
SDSS  J135636.71$-$001705.0	 & 	51942-0301-235	 & $-$27.80 & 5712 & 38 & 4.36 & 0.15 & $-$0.74 & 0.10 & & $-$21.40 & 5680 & 95 & 4.50 & 0.08 & $-$1.17 & 0.02\\
SDSS  J145319.68+010742.5	 & 	51994-0309-410	 & 152.40 & 5422 & 53 & 2.58 & 0.33 & $-$2.20 & 0.07 & & 148.40 & 5000 & 80 & 2.10 & 0.06 & $-$2.60 & 0.02\\
SDSS  J004029.17+160416.2	 & 	52342-0527-500	 & 50.10 & 5254 & 46 & 4.25 & 0.06 & $-$0.65 & 0.07 & & 52.40 & 5250 & 160 & 4.30 & 0.13 & $-$1.30 & 0.03\\
SDSS  J132832.61+020839.7	 & 	52435-0791-093	 & $-$5.10 & 6271 & 64 & 4.19 & 0.13 & $-$0.74 & 0.13 & & $-$5.50 & 5900 & 115 & 3.50 & 0.13 & $-$1.28 & 0.02\\
SDSS  J003602.17$-$104336.3	 & 	52378-0844-489	 & $-$48.70 & 5713 & 12 & 4.07 & 0.36 & $-$0.56 & 0.15 & & $-$47.20 & 6100 & 145 & 5.00 & 0.12 & $-$0.81 & 0.03\\
SDSS  J144705.99+555654.8	 & 	52764-1326-430	 & $-$0.50 & 5054 & 95 & 4.78 & 0.12 & $-$0.47 & 0.13 & & $-$0.00 & 5113 & 195 & 5.06 & 0.16 & $-$0.98 & 0.04\\
SDSS  J121821.60+053460.0	 & 	52786-1328-023	 & $-$27.00 & 5635 & 14 & 4.56 & 0.13 & $-$0.73 & 0.09 & & $-$26.70 & 5700 & 155 & 4.20 & 0.16 & $-$0.97 & 0.03\\
SDSS  J204227.48$-$002849.8	 & 	52786-1328-593	 & $-$54.20 & 5075 & 75 & 4.88 & 0.09 & $-$0.78 & 0.04 & & $-$50.60 & 5400 & 170 & 5.00 & 0.14 & $-$0.80 & 0.03\\
\tableline
\multicolumn{17}{c}{Subaru}  \\
\tableline
SDSS  J204101.22$-$002322.5	 & 	52146-0654-011	 & $-$150.20 & 6629 & 93 & 3.70 & 0.36 & $-$2.18 & 0.29 & & $-$146.63 & 6400 & 200 & 4.30 & 0.16 & $-$2.30 & 0.02\\
SDSS  J204728.84+001553.8	 & 	51817-0418-567	 & $-$44.60 & 6779 & 81 & 3.85 & 0.16 & $-$2.82 & 0.04 & & $-$50.05 & 6250 & 150 & 3.80 & 0.08 & $-$3.10 & 0.02\\
SDSS  J205322.46$-$000749.9	 & 	52466-0982-480	 & $-$419.20 & 6384 & 55 & 3.82 & 0.19 & $-$2.21 & 0.09 & & $-$418.39 & 6170 & 150 & 3.70 & 0.14 & $-$2.40 & 0.03\\
SDSS  J205458.93+004404.5	 & 	53289-1960-416	 & $-$66.60 & 5095 & 54 & 2.860 & 0.22 & $-$0.57 & 0.08 & & $-$67.70 & 5180 & 150 & 3.10 & 0.12 & $-$0.38 & 0.03\\
SDSS  J031249.63+001325.4	 & 	53401-2052-197	 & 10.30 & 6462 & 183 & 3.65 & 0.35 & $-$0.74 & 0.17 & & 13.53 & 6690 & 130 & 4.20 & 0.10 & $-$0.11 & 0.03\\
SDSS  J010531.72$-$002041.9	 & 	53534-2184-058	 & $-$366.10 & 6252 & 74 & 4.04 & 0.33 & $-$2.21 & 0.11 & & $-$371.12 & 6380 & 120 & 5.00 & 0.10 & $-$2.20 & 0.02\\
SDSS  J005227.41$-$002619.5	 & 	53534-2184-120	 & $-$337.10 & 5076 & 39 & 2.01 & 0.24 & $-$2.31 & 0.04 & & $-$334.18 & 5140 & 95 & 2.50 & 0.08 & $-$2.31 & 0.02\\
SDSS  J003802.72$-$001420.0	 & 	53534-2184-136	 & $-$85.40 & 5993 & 52 & 4.20 & 0.11 & $-$0.99 & 0.08 & & $-$86.53 & 6150 & 65 & 4.60 & 0.05 & $-$0.78 & 0.01\\
SDSS  J001652.51+001658.3	 & 	53713-2314-090	 & $-$273.80 & 6923 & 45 & 4.33 & 0.24 & $-$2.58 & 0.08 & & $-$272.56 & 6800 & 200 & 4.50 & 0.13 & $-$2.90 & 0.04\\
\tableline
\multicolumn{17}{c}{Keck-ESI}  \\
\tableline
SDSS  J234216.79$-$000603.1	 & 	52435-0981-085	 & 48.70 & 6747 & 82 & 3.13 & 0.29 & $-$2.03 & 0.16 & & 19.90 & 6856 & 144 & 3.25 & 0.14 & $-$2.00 & 0.08\\
SDSS  J205025.83$-$011103.8	 & 	52435-0981-123	 & 12.10 & \nodata & \nodata & \nodata & \nodata & \nodata & \nodata & & 48.32 & 4625 & 148 & 4.55 & 0.29 & $-$1.25 & 0.08\\
SDSS  J003159.54$-$001113.1	 & 	52443-0983-164	 & $-$22.20 & 4666 & 26 & 4.43 & 0.24 & $-$1.19 & 0.18 & & $-$23.34 & 4627 & 172 & 4.54 & 0.34 & $-$1.25 & 0.09\\
SDSS  J221855.69+000921.2	 & 	52442-0984-332	 & $-$48.30 & 5880 & 26 & 3.73 & 0.41 & $-$1.87 & 0.08 & & $-$39.81 & 5858 & 104 & 3.77 & 0.13 & $-$1.75 & 0.06\\
SDSS  J222116.19+005913.6	 & 	52589-1066-557	 & $-$39.10 & 5982 & 79 & 2.08 & 0.37 & $-$1.29 & 0.02 & & $-$45.09 & 5875 & 119 & 1.29 & 0.05 & $-$1.27 & 0.05\\
SDSS  J222725.18+003204.6	 & 	52523-1082-180	 & $-$31.10 & 5953 & 87 & 2.72 & 0.45 & $-$2.30 & 0.06 & & $-$34.82 & 6031 & 111 & 2.54 & 0.09 & $-$2.00 & 0.07\\
SDSS  J222542.47$-$003708.2	 & 	52591-1084-108	 & $-$210.10 & 6154 & 103 & 3.80 & 0.28 & $-$1.97 & 0.19 & & $-$198.35 & 6227 & 116 & 3.08 & 0.12 & $-$1.98 & 0.07\\
SDSS  J222005.05+011452.3	 & 	52531-1085-309	 & 2.10 & 4694 & 96 & 3.09 & 0.20 & $-$1.05 & 0.06 & & 70.08 & 4637 & 108 & 3.78 & 0.18 & $-$1.77 & 0.08\\
SDSS  J142409.29+533723.9	 & 	52929-1088-353	 & $-$228.90 & 6718 & 45 & 3.86 & 0.30 & $-$2.67 & 0.10 & & $-$240.26 & 6782 & 167 & 3.83 & 0.19 & $-$2.25 & 0.11\\
SDSS  J145758.20+504733.6	 & 	52591-1093-155	 & $-$183.90 & 6411 & 96 & 3.53 & 0.31 & $-$2.27 & 0.10 & & $-$194.14 & 6687 & 114 & 4.19 & 0.14 & $-$2.00 & 0.07\\
SDSS  J145543.59+510630.1	 & 	52932-1116-001	 & $-$12.10 & 5218 & 95 & 4.01 & 0.12 & $-$0.70 & 0.09 & & $-$33.26 & 5083 & 164 & 4.33 & 0.28 & $-$0.76 & 0.05\\
SDSS  J232541.95+001413.6	 & 	52993-1133-277	 & 194.10 & 5987 & 46 & 2.81 & 0.45 & $-$2.25 & 0.13 & & 207.14 & 6209 & 104 & 2.56 & 0.09 & $-$2.00 & 0.07\\
SDSS  J002140.87+004820.4	 & 	53228-1138-391	 & $-$107.70 & 5544 & 68 & 2.36 & 0.39 & $-$2.41 & 0.08 & & $-$102.12 & 5537 & 92 & 2.54 & 0.09 & $-$2.00 & 0.07\\
SDSS  J003828.39+003656.6	 & 	53228-1138-414	 & $-$95.10 & 5815 & 50 & 2.97 & 0.51 & $-$2.47 & 0.04 & & $-$101.10 & 6057 & 81 & 2.48 & 0.07 & $-$2.25 & 0.06\\
SDSS  J003928.61+010850.4	 & 	53228-1138-626	 & 39.30 & 5824 & 49 & 3.36 & 0.56 & $-$2.23 & 0.01 & & 39.54 & 5797 & 88 & 2.51 & 0.08 & $-$2.00 & 0.06\\
SDSS  J011135.53$-$002103.5	 & 	52592-1143-047	 & $-$137.80 & 6309 & 81 & 3.57 & 0.09 & $-$2.23 & 0.14 & & $-$171.93 & 6795 & 154 & 3.85 & 0.18 & $-$2.00 & 0.09\\
SDSS  J020100.13$-$004259.0	 & 	53238-1144-402	 & 5.40 & 4721 & 32 & 1.50 & 0.28 & $-$2.73 & 0.15 & & $-$3.70 & 5331 & 104 & 2.20 & 0.09 & $-$2.25 & 0.09\\
SDSS  J021748.78+002916.7	 & 	52992-1485-513	 & $-$83.90 & 6548 & 47 & 3.51 & 0.45 & $-$2.32 & 0.20 & & $-$86.04 & 6711 & 110 & 3.88 & 0.13 & $-$2.25 & 0.07\\
SDSS  J212748.24+003203.8	 & 	52932-1492-535	 & $-$54.20 & 5572 & 33 & 2.35 & 0.30 & $-$2.50 & 0.04 & & $-$57.35 & 5538 & 85 & 2.18 & 0.07 & $-$2.25 & 0.07\\
SDSS  J212935.16+121439.5	 & 	52937-1494-542	 & $-$333.50 & 5649 & 78 & 2.72 & 0.40 & $-$2.46 & 0.04 & & $-$328.87 & 5799 & 81 & 2.16 & 0.06 & $-$1.99 & 0.06\\
SDSS  J053234.96$-$003713.6	 & 	52944-1495-328	 & $-$92.40 & 6344 & 59 & 3.87 & 0.23 & $-$2.10 & 0.12 & & $-$92.83 & 6789 & 160 & 3.85 & 0.18 & $-$2.00 & 0.10\\
SDSS  J180922.45+223712.4	 & 	52914-1498-034	 & $-$11.10 & 6330 & 50 & 3.49 & 0.19 & $-$1.77 & 0.06 & & $-$14.50 & 6559 & 109 & 3.23 & 0.11 & $-$1.75 & 0.06\\
SDSS  J181001.41+230554.9	 & 	52941-1505-094	 & $-$48.00 & 4778 & 198 & 2.67 & 0.95 & $-$1.64 & 0.92 & & $-$46.50 & 4632 & 157 & 2.24 & 0.15 & $-$2.00 & 0.14\\
SDSS  J180728.56+223130.5	 & 	52944-1508-342	 & 88.50 & 6695 & 74 & 3.90 & 0.15 & $-$2.62 & 0.10 & & 75.54 & 7146 & 172 & 3.83 & 0.19 & $-$2.25 & 0.11\\
SDSS  J012617.95+060724.8	 & 	52945-1521-435	 & $-$181.50 & 5170 & 45 & 1.83 & 0.13 & $-$2.42 & 0.05 & & $-$196.13 & 5306 & 102 & 1.29 & 0.05 & $-$2.25 & 0.09\\
\enddata 
\label{params2}
\end{deluxetable} 

\end{landscape}

\begin{deluxetable}{lrrrrrrr}
\tablecolumns{8}
\tablewidth{0pc} 
\tablecaption{Parameters for the Standard Stars}
\tablehead{
\colhead{}	   &	\multicolumn{3}{c}{Literature}	&   \colhead{}   &
\multicolumn{3}{c}{HET}  \\
\cline{2-4} \cline{6-8} 	\\
\colhead{Star}         & 	
\colhead{$T_{\rm eff}$} & \colhead{$\log g$} & \colhead{ [Fe/H]} &  
\colhead{}	   &
\colhead{$T_{\rm eff}$} & \colhead{$\log g$} & \colhead{ [Fe/H]} }
\startdata
HD 8648	   & 	5790  & 4.28  &  0.13    & &  5833 &  4.36 &   0.09	\\
HD 71148   & 	5775  & 4.35  & $-0.03$  & &  5892 &  4.41 & $-0.04$	\\
HD 84737   & 	5906  & 4.22  &   0.12   & &  5929 &  4.12 &   0.07	\\
HD 84937   & 	6334  & 4.01  & $-2.11$  & &  6221 &  3.76 & $-2.07$
\enddata
\label{stds}
\end{deluxetable}

\begin{landscape}

\begin{deluxetable}{lrrrr}
\tablecolumns{5}
\tablewidth{0pc} 
\tablecaption{Comparison of Derived Stellar Atmospheric Parameters}
\tablehead{
\colhead{Analysis} & \colhead{Parameter}  &	\colhead{$<{\rm SSPP}-{\rm HI}>$}  &   
\colhead{$\sigma({\rm SSPP}-{\rm HI})$}   & \colhead{N}  }
\startdata
HET 	& Teff (K)        &    $3.11$\% &    2.75\%	&	81   \\

	& logg (dex)	  &   $0.08$	&    0.25	&	   \\
	
	& [Fe/H]  (dex)	  &    $-0.09$  &    0.12	&	   \\
	
	\tableline
	
OTHERS (Keck, Subaru)	& Teff (K)        &   $-0.58$\% &    3.14\%	 & 44 \\

	& logg (dex)	  &	$-0.03$	&    0.46	&	   \\
	
	& [Fe/H]  (dex)	  &    $-0.03$  &    0.41	&	   \\
		
\enddata
\label{compare}
\end{deluxetable}

\end{landscape}

\clearpage
\end{document}